%% file: DeepSICArxiv_v06.tex
\newif\ifsingle
\singletrue 

\newif\ifproofs

\ifsingle
\documentclass[12pt,draftclsnofoot, onecolumn]{IEEEtran}		
\else		
\documentclass[10pt,final, twocolumn]{IEEEtran}
\fi

\usepackage{times}
\usepackage{amsmath,dsfont}
\usepackage{amssymb,amsthm}
\usepackage{epsfig,verbatim}
\usepackage{setspace}
\usepackage{color}
\usepackage{cite}
\usepackage{epstopdf}
\usepackage{graphics}
\usepackage{accents}
\usepackage{acronym}
\usepackage[bookmarks,colorlinks]{hyperref}
\usepackage{booktabs}
\usepackage{mathtools}
\usepackage{algorithm}
\usepackage{algorithmic}
\usepackage{enumitem}
\usepackage{subcaption}

\input{Preamble.tex}

\ifsingle
\newcommand{\figWidth}{0.65\columnwidth}
\newcommand{\figSpace}{\vspace{-0.2cm}}
\newcommand{\includefig}[1]{\includegraphics[width = 0.65\columnwidth]{#1} 	\vspace{-0.2cm}}
\else
\newcommand{\figWidth}{\columnwidth}
\newcommand{\includefig}[1]{\includegraphics[width = 0.9\columnwidth]{#1} 	\vspace{-0.2cm}}
\newcommand{\figSpace}{\vspace{-0.6cm}}
\fi 

\IEEEoverridecommandlockouts 

\title{DeepSIC: Deep Soft Interference Cancellation for Multiuser MIMO Detection
}
\author{
	\IEEEauthorblockN{Nir Shlezinger,  Rong Fu, and Yonina C. Eldar
		\\
	} 	
\thanks{Parts of this work were presented in the 2020 IEEE International Conference on Acoustics, Speech, and Signal Processing (ICASSP), as the paper \cite{Shlezinger:20a}.
		This project has received funding from the Benoziyo Endowment Fund for the Advancement of Science, the	Estate of Olga Klein – Astrachan, the European Union’s Horizon 2020 research and innovation program under grant No. 646804-ERC-COG-BNYQ, and from the Israel Science Foundation under grant No. 0100101.
		N. Shlezinger  and Y. C. Eldar are with the Faculty of Math and CS, Weizmann Institute of Science, Rehovot, Israel (e-mail: nirshlezinger1@gmail.com; yonina@weizmann.ac.il). 	
			R. Fu is with the Department of EE,  Tsinghua University, Beijing, China (e-mail: fu-r16@mails.tsinghua.edu.cn).
	}

	\vspace{-1.0cm}
	
}
\vspace{-0.75cm}

\begin{document}
	
	\maketitle
	\pagestyle{plain}
	\thispagestyle{plain}
	\begin{abstract}
	Digital receivers are required to recover the transmitted symbols from their observed channel output. In multiuser multiple-input multiple-output (MIMO) setups, where multiple symbols are simultaneously transmitted, accurate symbol detection is challenging. A family of algorithms capable of reliably recovering multiple symbols is based on interference cancellation. However, these methods assume that the channel is linear, a model which does not reflect many relevant channels, as well as require accurate channel state information (CSI), which may not be available. 
	In this work we propose a multiuser MIMO receiver which learns to jointly detect in a data-driven fashion, without assuming a specific channel model or requiring CSI. In particular, we propose a data-driven implementation of  the iterative soft interference cancellation (SIC) algorithm which we refer to as DeepSIC. The resulting symbol detector is based on integrating dedicated machine-learning methods into the iterative SIC algorithm. DeepSIC learns to carry out joint detection from a limited set of training samples without requiring the channel to be linear and its parameters to be known. 
	Our numerical evaluations demonstrate that for linear channels with full CSI, DeepSIC approaches the performance of iterative SIC, which is comparable to the optimal performance, and outperforms previously proposed learning-based MIMO receivers. Furthermore, in the presence of CSI uncertainty, DeepSIC significantly outperforms model-based approaches. Finally, we show that DeepSIC accurately detects symbols in non-linear channels, where conventional iterative SIC fails even when accurate CSI is available.  
	\end{abstract}

	\vspace{-0.4cm}
	\section{Introduction}
	\vspace{-0.1cm}
	Modern communications systems are subject to constantly growing throughput requirements. In order to meet these demands, receivers are commonly equipped with multiple antennas, and communicate with several transmitters simultaneously to increase the spectral efficiency \cite{Foschini:98}. Such scenarios, referred to as  multiuser \ac{mimo} networks,   are typically encountered in uplink cellular systems, where the number of transmitters as well as receiver antennas can be very large, as in, e.g., massive \ac{mimo} communications \cite{Marzetta:10}.  
	
	One of the main challenges in multiuser \ac{mimo} systems is symbol detection, namely, the recovery of the multiple transmitted symbols at the receiver.  Conventional detection algorithms, such as those based on the \ac{map} rule which jointly recovers all the symbols simultaneously, become infeasible as the number of symbols grows. Alternatively, low complexity separate detection, in which each symbol is recovered individually while treating the rest of the symbols, i.e., the interference, as noise, is strictly sub-optimal \cite[Ch. 6]{ElGamal:11}, and thus results in degraded throughput \cite{Shlezinger:17}.  An attractive approach, both in terms of complexity and in performance, is interference cancellation \cite{Andrews:05}. This family of  algorithms implement separate detection, either successively or in parallel, and uses the estimates to facilitate the recovery of the remaining symbols, essentially trading complexity for detection delay.  While these methods are prone to error propagation, its effect can be dramatically mitigated by using soft symbol estimates \cite{Choi:00,Wang:99,Alexander:99}, achieving near \ac{map} performance with controllable complexity.    
	
	\label{txt:NonLinear}
	\textcolor{NewColor}{
	These aforementioned detection strategies are model-based, namely, they require complete knowledge of the channel model as well as its parameters.  When the channel model is unknown, highly complex, or does not faithfully represent the  physical environment, these methods cannot be applied. Furthermore, common model-based detection techniques, including interference cancellation schemes, typically assume linear Gaussian channels, in which the noise obeys a Gaussian distribution and the effect of the interference is additive and can thus be canceled by subtraction. Many important future wireless communication scenarios, involving, e.g.,  quantization-constrained receivers \cite{Studer:16,Shlezinger:19b}, transmission with non-linear amplifiers \cite{Iofedov:15}, and communication in the presence of radar interference \cite{Zheng:19}, do not obey the linear Gaussian model. Furthermore, various other communication systems, such as optical networks \cite{Khalighi:14}, power-line communications \cite{Shlezinger:18b}, and molecular communications \cite{Farsad:17}, cannot be accurately modeled as linear Gaussian channels. Consequently,  the applicability of model-based interference cancellation methods is limited. In addition, even when the channel model is linear and known, inaccurate knowledge of the parameters of the channel, namely,  \ac{csi} uncertainty, can significantly degrade the performance of model-based detection mechanisms. This motivates the study of data-driven model-agnostic detection methods. }
	
	An alternative to model-based detection algorithms, which is gaining considerable interest recently, is to utilize \acl{ml} tools. Over the last decade, \acl{ml} based systems, and particularly \acp{dnn}, have revolutionized numerous research areas, including computer vision and speech processing \cite{LeCun:15}. \acl{ml} schemes are gradually influencing the design of digital communication systems, resulting in a multitude of recent works on the application of \acp{dnn} in communications; see detailed surveys in  \cite{Oshea:17, Simeone:18, Mao:18, Gunduz:19, He:19c,Balatsoukas:19}. Unlike model-based receivers, which implement a specified detection rule, \acl{ml} based receivers learn how to map the channel outputs into the transmitted symbols from training, namely, they operate in a data-driven manner, and are typically capable of disentangling the semantic information in complex environments \cite{Bengio:09}. Furthermore, once trained, \ac{dnn}-based receivers can implement complicated computations with affordable complexity, making them a promising approach to implement \ac{mimo} detection. 
	
	Broadly speaking, previously proposed \acl{ml} based receivers can be divided into two main categories: Conventional \acp{dnn} and unfolded networks.  The first group replaces the receiver processing with a \ac{dnn} architecture which is established in the \acl{ml} literature. A-priori knowledge of the channel model is accounted for in the selection of the network type, which is typically treated as a black box. For example, \acp{rnn} were applied for  decoding sequential codes in \cite{Kim:18}; the works \cite{Farsad:18,Liao:19} used sliding bi-directional \acp{rnn} for \ac{isi} channels with long memory; reservoir computing was proposed for recovering distorted \ac{ofdm} signals in \cite{Mosleh:18}; and the work \cite{Caciularu:18} used variational autoencoders for unsupervised equalization. Such \acp{dnn},  which use conventional network architectures that are ignorant of the underlying channel model, can typically operate reliably in various scenarios with or without \ac{csi} and channel model knowledge \cite{Ye:18}, assuming that they were properly trained for the specific setup. Nonetheless, black box \acp{dnn} tend to have a large number of parameters, and thus require a large number of samples to train \cite{Balatsoukas:19}, limiting their application in dynamic environments, which are commonly encountered in communications. 
	
	Unlike conventional \acp{dnn}, which utilize established architectures, in unfolded receivers the network structure is designed following a model-based algorithm.  In particular, deep unfolding is a method for converting an iterative algorithm into a \ac{dnn} by designing each layer of the network to resemble a single iteration \cite{Hershey:14}. The resulting \ac{dnn} tends to demonstrate improved convergence speed and robustness compared to the model-based algorithm \cite{Gregor:10,Solomon:18,Li:19}. In the context of \ac{mimo} symbol detection, the works \cite{Wiesel:19,Corlay:18, Takabe:19} designed deep receivers by unfolding the projected gradient descent algorithm for recovering the \ac{map} solution, and \cite{Khobahi:19} proposed to recover continuous-valued signals obtained from one-bit quantized measurements by unfolding gradient descent optimization. 
	\label{txt:NewRefs}
	\textcolor{NewColor}{
	Iterative message passing algorithms, which are known to facilitate multi-user \ac{mimo} detection and decoding at controllable complexity \cite{Liu:16,Liu:1b9}, were used as a basis for designing data-driven \ac{mimo} detectors in  \cite{Guo:19, He:19a,He:19b} as well as for channel estimation and user activity detection in \cite{Zhang:19}.}
	 Compared to conventional \acp{dnn}, unfolded networks are typically interpretable, and tend to have a smaller number of parameters, and can thus be trained quicker \cite{Balatsoukas:19}. However,  these previously proposed receivers all assume a linear channel with Gaussian noise, in which \ac{csi} is either available \cite{Wiesel:19,Corlay:18, Takabe:19, He:19a,Khobahi:19} or estimated from pilots \cite{He:19b}.  Consequently, these methods thus do not capture the potential of \acl{ml} in being model independent, and are applicable only under specific channel setups.  
	
	In our previous work \cite{Shlezinger:19a} we proposed ViterbiNet, which is a data-driven implementation of the Viterbi algorithm for detecting symbols transmitted over channels with finite memory. Instead of implementing the receiver as a conventional \ac{dnn}, or alternatively, unfolding the Viterbi algorithm, we replaced its model-based computations with simple dedicated \acp{dnn}. The resulting receiver was thus capable of implementing Viterbi detection in a data-driven fashion using a relatively small number of parameters, while being channel-model-independent. 
	Since ViterbiNet is a data-driven implementation of the Viterbi algorithm whose computational complexity grows rapidly with the cardinality of the channel input,  it is not suitable in its current form for \ac{mimo} detection. However, the fact that it is capable of learning to implement symbol detection  from small training sets motivates the design of a data-driven model-ignorant \ac{mimo} symbol detector by integrating \acl{ml} into an established detection algorithm, which is the focus of the current work.

	Here, we design a data-driven \ac{mimo} detector which is based on model-based interference cancellation methods while being channel-model-independent. In particular, we base our approach on the iterative \ac{sic} symbol detection algorithm proposed in \cite{Choi:00} as a model-based method which is capable of approaching \ac{map} performance at affordable complexity. Then, we propose Deep\ac{sic}, in which the model-based building blocks of iterative \ac{sic} are replaced with simple dedicated \acp{dnn}. Deep\ac{sic} thus implements iterative \ac{sic} in a data-driven fashion,  learning to implement \ac{map}-comparable symbol detection from a limited training set. Furthermore, the fact that Deep\ac{sic} learns to cancel the interference from training allows it to operate in non-linear setups, where model-based iterative \ac{sic}, which assume that the interference is additive and can be canceled by subtracting it, is not applicable. 
	
	We propose two methods for training Deep\ac{sic}. First we discuss how to jointly train the building blocks consisting of the iterations of iterative \ac{sic} in an end-to-end manner. Then, we show how the same training set can be used to train different subsets of the overall network sequentially, by exploiting our prior knowledge of the specific role of each block in the iterative \ac{sic} algorithm. The resulting sequential training method allows the receiver to learn its symbol detection mapping from a smaller number of samples. The ability to accurately train with a small training set facilitates its applicability for online training, making the sequential approach attractive for dynamic environments in which the receiver has to frequently retrain by exploiting the inherent redundancy induced by digital communication protocols   \cite{Shlezinger:19a}.  
	
	We numerically demonstrate the benefits of Deep\ac{sic} in a simulation study. We show that it is capable of approaching the performance of the model-based iterative \ac{sic} algorithm in standard linear \ac{mimo} channels with Gaussian noise, and that it achieves improved error rate performance compared to previously proposed data-driven \ac{mimo} receivers. We also observe that, in the presence of \ac{csi} uncertainty, the performance of the model-based iterative \ac{sic}, as well as the \ac{map} detector, is significantly degraded, while   Deep\ac{sic}, which exploits the generalization capabilities of \acp{dnn}, is  capable of reliably detecting the transmitted symbols.  Then, we consider non-linear channels, where Deep\ac{sic} is shown to continue to achieve \ac{map} comparable performance, while substantially outperforming iterative \ac{sic}, which is incapable of canceling the non-additive interference.  \textcolor{NewColor}{Finally, we show how the ability of DeepSIC to train with small training sets can facilitate online tracking of dynamic channels in a self-supervised manner.} Our results demonstrate that efficient and robust communication systems can be realized by properly integrating \acl{ml} methods into model-based algorithms.

 	The rest of this paper is organized as follows: In Section~\ref{sec:Preliminaries} we present our system model and review the iterative \ac{sic} algorithm. Section~\ref{sec:Net}  proposes Deep\ac{sic}, which is a receiver architecture integrating \acp{dnn} into the  iterative \ac{sic} method. 
 	Section~\ref{sec:Sims} details numerical training and performance results of the proposed receiver, and Section~\ref{sec:Conclusions} provides concluding remarks.

 	\smallskip
 	Throughout the paper, we use upper-case letters for \acp{rv}, e.g. $X$, and calligraphic letter for sets, for example, $\mySet{X}$.
 	Boldface lower-case letters denote vectors, e.g., ${\myVec{x}}$ is a deterministic vector, and $\myVec{X}$ is a random vector, and
 	the $i$th element of ${\myVec{x}}$ is written as $\left( {\myVec{x}}\right) _i$. 
 	\textcolor{NewColor}{Since upper-case boldface letters are reserved for random vectors, we use upper-case Sans-Sarif fonts for matrices as in  \cite{Shlezinger:17b, Shlezinger:18b, Shlezinger:19a, Shlezinger:19b}, e.g., $\myMat{X}$ is a deterministic matrix.}  	
 	The probability measure of an \ac{rv} $X$ evaluated at $x$ is denoted $\Pdf{X}(x)$,  
 	%
 	%
 	 $\mySet{R}$ is the set of real numbers, and $(\cdot)^T$ is the transpose operator. 
 	All logarithms are taken to basis 2. 
	\vspace{-0.2cm}
	\section{System Model and Iterative \ac{sic}}
	\label{sec:Preliminaries}
	\vspace{-0.1cm}
	
	\subsection{System Model}
	\label{subsec:Model}
\begin{figure}
	\centering
	\includefig{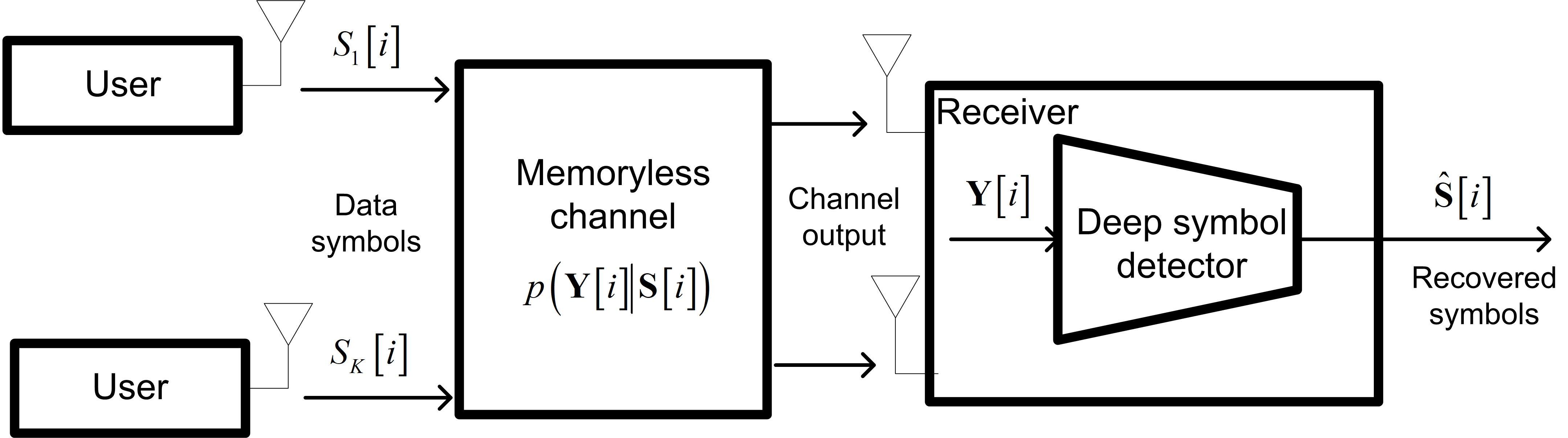} 
	\caption{System model.}
	\label{fig:BasicModel2}
\end{figure}

	\vspace{-0.1cm}
	We consider an uplink  system in which $\Nusers$ single antenna users communicate with a receiver equipped with $\Nantennas$ antennas over a memoryless stationary channel.  
	At each time instance $i$, the $k$th user, $k \in \{1,2,\ldots, \Nusers\}\triangleq \NusersSet$, transmits a symbol $S_k[i]$  drawn from a constellation $\mySet{S}$ of size $\CnstSize$, i.e., $| \mySet{S}| = \CnstSize$. Each symbol is uniformly distributed over $\mySet{S}$, and the symbols transmitted by different users are mutually independent.   We use $\myY[i]\in \mySet{R}^{\Nantennas}$ to denote the channel output at time index $i$. 
	\label{txt:Complex}
	While we focus on real-valued channels, the system model can be adapted to complex-valued channels, as complex vectors can be equivalently represented using real vectors of extended dimensions. 
	\textcolor{NewColor}{
	 Accordingly, we do not restrict the constellation set $\mySet{S}$ to take real values, and  the receiver architectures detailed in the sequel, which are formulated for real-valued systems, can be applied in complex channels.} 
	 Since the channel is memoryless, $\myY[i] $ is given by some stochastic mapping of $\myS[i] \triangleq \big[S_1[i], S_2[i], \ldots, S_{\Nusers}[i]\big]^T$, represented by the conditional distribution measure $\Pdf{\myY[i] | \myS[i]} (\cdot | \cdot)$.  The fact that the channel is stationary implies that  this conditional distribution does not depend on the  index $i$, and is thus denoted henceforth by $\Pdf{\myY | \myS} (\cdot | \cdot)$. An illustration of the system is depicted in Fig. \ref{fig:BasicModel2}.

	We focus on the problem of recovering the transmitted symbols  $\myS[i]$ from the channel output $\myY[i]$. 
	The optimal detection rule which minimizes the probability of error given a channel output realization $\myY[i] = \myVec{y}$ is the \ac{map} detector. Letting $\Pdf{\myS | \myY} (\cdot | \cdot)$ be the conditional distribution of $\myS[i]$ given $\myY[i]$, the \ac{map} rule is given by
	\begin{equation}
	\label{eqn:MAP}
	\hat{\myVec{s}}_{\rm MAP}[i] \triangleq \mathop{\arg \max}\limits_{\myVec{s} \in \mySet{S}^{\Nusers}} \Pdf{\myS | \myY} (\myVec{s} | \myVec{y}).
	\end{equation}
	
	The \ac{map} detector jointly recovers the symbols of all users  by searching over a set of $\CnstSize^{\Nusers}$ different possible input combinations, and thus becomes infeasible when the number of users $\Nusers$ grows. For example, when binary constellations are used, i.e., $\CnstSize = 2$, the number of different channel inputs is larger than $10^6$ for merely $\Nusers = 20$ users. 
	Furthermore, the \ac{map} detector requires accurate knowledge of the channel model, i.e., the conditional distribution $\Pdf{\myY | \myS} (\cdot | \cdot)$ must be fully known.  
	 A common strategy to implement joint detection with affordable computational complexity, suitable for channels in which $\myY[i]$ is given by a linear transformation of $\myS[i]$ corrupted by additive noise, is interference cancellation \cite{Andrews:05}.  Interference cancellation refers to a family of algorithms which implement joint detection in an iterative fashion by recovering a subset of $\myS[i]$ based on the channel output as well as an estimate of the remaining interfering symbols. These algorithms facilitate the recovery of the subset of $\myS[i]$ from the channel output by canceling the effect of the estimated interference using knowledge of the channel parameters, and specifically, how each interfering symbol contributes to the channel output.

	Our goal is to design a data-driven method for recovering  $\myS[i]$ from the channel output $\myY[i]$, \textcolor{NewColor}{which learns its detection mapping using a training set of $\Ntraining$ pairs of realizations of $\myY[i]$ and corresponding  $\myS[i]$, denoted $\{\tilde{\myVec{s}}_j,\tilde{\myVec{y}}_j\}_{j=1}^{\Ntraining}$.} 
	In particular, in our model the receiver  knows the constellation $\mySet{S}$, and that the channel is stationary and memoryless. We do not assume that the channel is linear nor  that the receiver knows the conditional probability measure $\Pdf{\myY | \myS} (\cdot | \cdot)$.  
	Following the approach of \cite{Shlezinger:19a,Shlezinger:20b, Shlezinger:20c}, we design our network to implement interference cancellation in a data-driven fashion. In particular, our proposed receiver is based on the iterative  \ac{sic} algorithm proposed in \cite{Choi:00}. 
	Therefore, as a preliminary step to designing the data-driven detector, we review iterative \ac{sic} in the following subsection.

	\vspace{-0.2cm}
	\subsection{Iterative Soft Interference Cancellation}
	\label{subsec:SIC}
\begin{figure}
		\centering
		\includegraphics[width = \columnwidth]{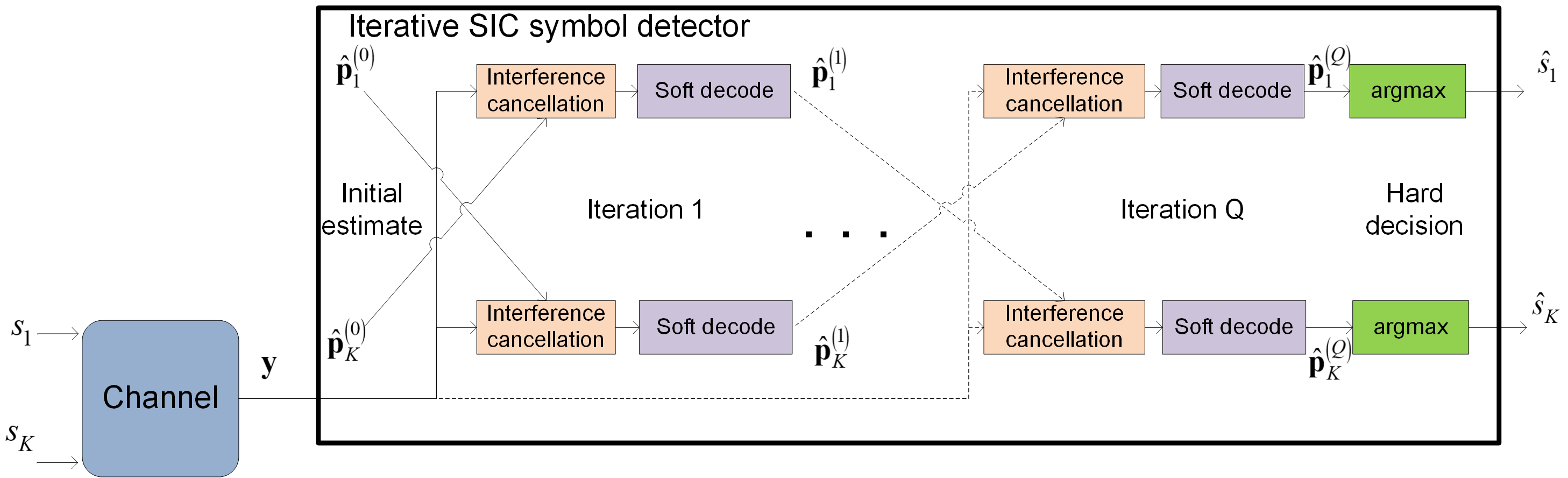} 
		\vspace{-0.2cm}
		\caption{Soft iterative interference cancellation illustration.}
		\label{fig:SoftIC1}
	\end{figure}

	\vspace{-0.1cm}
	The iterative  \ac{sic} algorithm proposed in \cite{Choi:00} is a multiuser  detection method which combines multi-stage (parallel) interference cancellation \cite{Varanasi:90} with soft decisions.  
	Broadly speaking, the detection method operates in an iterative fashion, where in each iteration, an estimate of the conditional distribution of $S_k[i]$ given the channel output $\myY[i] = \myVec{y}$ is generated for every user $k \in \NusersSet$ using the corresponding estimates of the interfering symbols $\{S_l[i]\}_{l \neq k}$ obtained in the previous iteration. By repeating this procedure iteratively, the conditional distribution estimates are refined, allowing to accurately recover each symbol from the output of the last layer using hard decision.  This iterative procedure is illustrated in Fig. \ref{fig:SoftIC1}.  
	
	To formulate the algorithm, we consider a channel whose output is obtained as a linear transformation of its input corrupted by \ac{awgn}, i.e., 
	\begin{equation}
	\label{eqn:Linear channel}
	\myY[i] = \myMat{H} \myS[i] + \myVec{W}[i],
	\end{equation}
	where $\myMat{H} \in \mySet{R}^{\Nantennas \times \Nusers}$ is an a-priori known channel matrix, and $\myVec{W}[i] \in \mySet{R}^{\Nantennas}$ is a zero-mean multivariate Gaussian vector with covariance $\SigW \myI_{\Nusers}$, independent of $\myS[i]$.
	
	Iterative soft interference cancellation consists of $\Niter$ iterations, where  each iteration indexed  $q \in \{1,2,\ldots, \Niter\} \triangleq \NiterSet$ generates $\Nusers$ distribution vectors $\hat{\myVec{p}}_k^{(q)} \in \mySet{R}^{\CnstSize}$, $k \in \NusersSet$. These vectors are computed from the channel output $\myVec{y}$ as well as the distribution vectors obtained at the previous iteration, $\{ \hat{\myVec{p}}_k^{(q-1)}\}_{k=1}^{\Nusers}$, as detailed in the sequel. The entries of  $\hat{\myVec{p}}_k^{(q)}$ are estimates of the distribution of $S_k[i]$ for each possible symbol in $\mathcal{S}$, given the channel output $\myY[i] = \myVec{y}$ and assuming that the interfering symbols $\{S_l[i]\}_{l \neq k}$ are distributed via $\{ \hat{\myVec{p}}_l^{(q-1)}\}_{l \neq k}$. Note that for binary constellations, i.e., $\CnstSize = 2$, $\hat{\myVec{p}}_k^{(q)}$ can be represented using a single scalar value, as $\big(\hat{\myVec{p}}_k^{(q)}\big)_2 = 1- \big(\hat{\myVec{p}}_k^{(q)}\big)_1$. 
	
	Every iteration consists of two steps, carried out in parallel for each user: {\em Interference cancellation}, and {\em soft detection}. 
	Focusing on the $k$th user and the $q$th iteration, the interference cancellation stage first computes the expected values and variances of $\{S_l[i]\}_{l \neq k}$ based on  $\{ \hat{\myVec{p}}_l^{(q-1)}\}_{l \neq k}$. Letting $\{\alpha_m\}_{m=1}^{\CnstSize}$ be the indexed elements of the constellation set $\mySet{S}$, the expected values and variances are computed via
	\begin{equation}
	\label{eqn:ExpVal1}
	e_l^{(q-1)} = \sum\limits_{\alpha_m \in \mySet{S}} \alpha_m \left( \hat{\myVec{p}}_l^{(q-1)}\right)_m, 
	\end{equation}
	and 
	\begin{equation}
	\label{eqn:ExpVar1}
	v_l^{(q-1)} = \sum\limits_{\alpha_m \in \mySet{S}}\left(  \alpha_m - e_l^{(q-1)}\right) ^2 \left( \hat{\myVec{p}}_l^{(q-1)}\right)_m, 
	\end{equation}
	respectively. 
	The contribution of the interfering symbols from $\myVec{y}$ is then canceled by replacing these symbols with $\{e_l^{(q-1)}\}$ and subtracting their contribution from the channel output. By letting $\myVec{h}_l$ denote the $l$th column of $\myMat{H}$, the interference canceled channel output is given by
	\begin{subequations}
	\begin{align}
	\label{eqn:Cancelled1} 
	\myVec{Z}_k^{(q)}[i] 
	&= \myVec{Y}[i] - \sum\limits_{l \neq k} \myVec{h}_l e_l^{(q-1)}  \\
	&= \myVec{h}_k S_k[i] + \sum\limits_{l \neq k} \myVec{h}_l (S_l[i]- e_l^{(q-1)}) + \myVec{W}[i].
	\label{eqn:Cancelled1a} 
	\end{align} 
\end{subequations}
	Substituting the channel output $\myVec{y}$ into \eqref{eqn:Cancelled1}, the realization of the interference canceled $\myVec{Z}_k^{(q)}[i]$, denoted  $\myVec{z}_k^{(q)}$, is obtained. 
	
	To implement soft detection, it is assumed that $\tilde{W}_k^{(q)}[i] \triangleq \sum\limits_{l \neq k} \myVec{h}_l (S_l[i]- e_l^{(q-1)}) + \myVec{W}[i]$ obeys a zero-mean Gaussian distribution, independent of $S_k[i]$, and that its covariance is given by
	\begin{equation}
	\CovMat{\tilde{W}_k^{(q)}} = \SigW \myI_{\Nusers} + \sum\limits_{l \neq k}v_l^{(q-1)}  \myVec{h}_l \myVec{h}_l^T.
	\end{equation}
	Combining this assumption with \eqref{eqn:Cancelled1a}, the conditional distribution of $\myVec{Z}_k^{(q)}$ given $S_k[i] = \alpha_m$ is multivariate Gaussian with mean value  $\myVec{h}_k \alpha_m$ and covariance $\CovMat{\tilde{W}_k^{(q)}}$. Since $\myVec{Z}_k^{(q)}[i] $ is given by a bijective transformation of $\myVec{Y}[i]$, it holds that $\Pdf{S_k | \myY}(\alpha_m | \myVec{y}) = \Pdf{S_k |  \myVec{Z}_k^{(q)}}(\alpha_m | \myVec{z}_k^{(q)})$ for each $\alpha_m \in \mySet{S}$ under the above assumptions. Consequently, the conditional distribution of $S_k[i]$ given $\myY[i]$ is approximated from the conditional distribution of $\myVec{Z}_k^{(q)}$ given $S_k[i]$ via Bayes theorem. Since the symbols are equiprobable, this estimated conditional distribution is computed as
	\ifsingle
	\begin{align}
	\left( \hat{\myVec{p}}_k^{(q)}\right)_m 
	&= \frac{\Pdf{ \myVec{Z}_k^{(q)} | S_k}( \myVec{z}_k^{(q)} | \alpha_m)}{\sum\limits_{\alpha_{m'}\in\mySet{S} }\Pdf{ \myVec{Z}_k^{(q)} | S_k}( \myVec{z}_k^{(q)} | \alpha_{m'})} \notag \\
	&= \frac{\exp \left\{-\frac{1}{2} \left( \myVec{z}_k^{(q)} - \myVec{h}_k\alpha_m \right)^T\CovMat{\tilde{W}_k^{(q)}}^{-1} \left( \myVec{z}_k^{(q)} - \myVec{h}_k\alpha_m \right)   \right\} } { \sum\limits_{\alpha_{m'}\in\mySet{S} }\exp \left\{-\frac{1}{2} \left( \myVec{z}_k^{(q)} - \myVec{h}_k\alpha_{m'} \right)^T\CovMat{\tilde{W}_k^{(q)}}^{-1} \left( \myVec{z}_k^{(q)} - \myVec{h}_k\alpha_{m'} \right)   \right\} }.
	\label{eqn:CondDist2}
	\end{align}
	\else
		\begin{align}
	&\left( \hat{\myVec{p}}_k^{(q)}\right)_m 
	= \frac{\Pdf{ \myVec{Z}_k^{(q)} | S_k}( \myVec{z}_k^{(q)} | \alpha_m)}{\sum\limits_{\alpha_{m'}\in\mySet{S} }\Pdf{ \myVec{Z}_k^{(q)} | S_k}( \myVec{z}_k^{(q)} | \alpha_{m'})} \notag \\
	&= \!\frac{\exp \left\{\!-\frac{1}{2} \left( \myVec{z}_k^{(q)} \!- \!\myVec{h}_k\alpha_m \right)^T\CovMat{\tilde{W}_k^{(q)}}^{-1} \left( \myVec{z}_k^{(q)} \!-\! \myVec{h}_k\alpha_m \right)   \right\} } { \sum\limits_{\alpha_{m'}\in\mySet{S} }\!\exp \left\{\!-\frac{1}{2} \left( \myVec{z}_k^{(q)} \!-\! \myVec{h}_k\alpha_{m'} \right)^T\CovMat{\tilde{W}_k^{(q)}}^{-1} \left( \myVec{z}_k^{(q)} \!-\! \myVec{h}_k\alpha_{m'} \right)   \right\} }.
	\label{eqn:CondDist2}
	\end{align}
	\fi

	After the final iteration, the symbols are detected by taking the symbol which maximizes the estimated conditional distribution for each user, i.e., 
	\begin{equation}
	\label{eqn:HardDet}
	\hat{s}_k = \mathop{\arg \max}\limits_{m \in \{1,\ldots,\CnstSize\}}\left( \hat{\myVec{p}}_k^{(\Niter)}\right)_m.  
	\end{equation}
	The overall joint detection scheme is summarized below as Algorithm \ref{alg:Algo1}.
	 The initial estimates $\{\hat{\myVec{p}}_k^{(0)}\}_{k=1}^{\Nusers}$ can be arbitrarily set. For example, these may be chosen based on a linear separate estimation of each symbol for $\myVec{y}$, as proposed in \cite{Choi:00}. 

\begin{algorithm}
	\caption{Iterative Soft Interference Cancellation Algorithm}
	\label{alg:Algo1}
	\begin{algorithmic}[1]
		\STATE \underline{Input}: Channel output $\myVec{y}$.
		\STATE \underline{Initialization}: Set $q=1$, and generate an initial guess of the conditional distributions $\{\hat{\myVec{p}}_k^{(0)}\}_{k =1}^{\Nusers}$. 
		\STATE \label{stp:MF1} Compute the expected values $\{e_k^{(q-1)}\}$ and variances $\{v_k^{q-1}\}$ via \eqref{eqn:ExpVal1}-\eqref{eqn:ExpVar1}, respectively.
		\STATE \label{stp:IC} {\em Interference cancellation:} For each $k \in \NusersSet$ compute $\myVec{z}_k^{(q)}$ via \eqref{eqn:Cancelled1}.
		\STATE \label{stp:SoftDec} {\em soft detection:} For each $k \in \NusersSet$, estimate the conditional distribution $\hat{\myVec{p}}_k^{(q)}$ via \eqref{eqn:CondDist2}.
		\STATE Set $q := q+1$. If $q \le \Niter $ go to Step \ref{stp:MF1}.	
		
		\STATE  \underline{Output}: Hard detected output $\hat{\myVec{s}}$, obtained via \eqref{eqn:HardDet}.
	\end{algorithmic}
\end{algorithm}

\vspace{-0.2cm}
\subsection{Advantages and Challenges of Iterative \ac{sic}}
\label{subsec:SICProsCons}
\vspace{-0.1cm}
Iterative \ac{sic} has several notable advantages as a joint detection method: In terms of computational complexity, it replaces the joint exhaustive search over all different channel input combinations, required by the \ac{map} detector \eqref{eqn:MAP}, with a set of computations carried out separately for each user. Hence, its computational complexity only grows linearly with the number of users \cite{Andrews:05}, making it feasible also with large values of $\Nusers$.  Unlike conventional separate detection, in which the symbol of each user is recovered individually while treating the interference as noise, the iterative procedure refines the separate estimates sequentially, and the usage of soft values mitigates the effect of error propagation. Consequently, Algorithm \ref{alg:Algo1} is capable of achieving performance approaching that of the \ac{map} detector, which is only feasible for small values of $\Nusers$. 
The iterative process trades computational complexity for increased detection delay. However, it is numerically observed in \cite{Choi:00} that a relatively small number of iterations, such as $\Niter = 5$, are sufficient for achieving substantial performance gains over separate detection with only a small additional detection delay.   

Iterative \ac{sic} is specifically designed for linear channels of the form \eqref{eqn:Linear channel}. In particular, the interference cancellation in Step \ref{stp:IC} of Algorithm \ref{alg:Algo1} requires the contribution of the interfering symbols to be additive. This limits the application of the algorithm in non-linear channels, such as those encountered in optical communications \cite{Khalighi:14}, or, alternatively, the models arising in the presence of low-resolution quantizers \cite{Studer:16}  and non-linear power amplifiers \cite{Iofedov:15}. Additionally, the fact that the distribution of the interference canceled channel output $\myVec{Z}_k^{(q)}$ given $S_k[i]$ is approximated as Gaussian in Algorithm \ref{alg:Algo1} may  degrade the performance in channels which do not obey the linear Gaussian model \eqref{eqn:Linear channel}.

Furthermore, even when the channel obeys the linear model of \eqref{eqn:Linear channel}, iterative \ac{sic} requires full \ac{csi}, i.e.,  knowledge of the channel matrix $\myMat{H}$ and the noise variance $\SigW$. Acquiring such knowledge may entail substantial overhead. It is in fact crucial for the receiver to have accurate \ac{csi}, since the performance of Algorithm \ref{alg:Algo1} is heavily degraded in the presence of  \ac{csi} errors, as empirically demonstrated in Section \ref{sec:Sims}. 

The dependence  on accurate \ac{csi} and the assumption of linear channels are not unique to iterative \ac{sic}, and are in fact common to most interference cancellation based joint detection algorithms \cite{Andrews:05}. These limitations motivate the design of a joint detector which exploits the computational feasibility of interference cancellation methods  while operating in a data-driven fashion. We specifically select iterative \ac{sic} since it is  capable of achieving \ac{map}-comparable performance, with a structure that can be readily converted to be data-driven. This is a result of the fact that its specific model-based computations, i.e., Steps \ref{stp:IC}-\ref{stp:SoftDec} in Algorithm \ref{alg:Algo1}, can be naturally implemented using relatively simple \acl{ml} methods.  The resulting receiver, detailed in the following section, integrates \acl{ml} methods into Algorithm \ref{alg:Algo1}, allowing it to be implemented for arbitrary memoryless stationary channels without requiring a-priori knowledge of the channel model and its parameters.

\vspace{-0.2cm}
\section{Deep\ac{sic}}
\label{sec:Net}
\vspace{-0.1cm} 
In this section, we present a data-driven implementation of iterative \ac{sic}. 
To formulate the proposed receiver, we first derive the \acl{ml} based receiver architecture, which we call Deep\ac{sic}, in Subsection \ref{subsec:Derivation}. Then, we present methods for training  the \acp{dnn} embedded in the receiver in Subsection \ref{subsec:Training}, and discuss its pros and cons in Subsection \ref{subsec:Discussion}. 

\subsection{Data-Driven Receiver Architecture}
\label{subsec:Derivation}
\vspace{-0.1cm}
Here, we present a receiver architecture which implements iterative \ac{sic} in a data-driven fashion. Following the approach of \cite{Shlezinger:19a, Shlezinger:20b}, we wish to keep the overall structure of the iterative \ac{sic} algorithm, depicted in Fig. \ref{fig:SoftIC1}, while replacing the channel-model-based computations with dedicated suitable \acp{dnn}. To that aim, we note that  iterative \ac{sic}  can be viewed as a set of interconnected basic building blocks, each implementing the two stages of interference cancellation and soft detection, i.e., Steps \ref{stp:IC}-\ref{stp:SoftDec} of Algorithm \ref{alg:Algo1}. While the high level architecture of Fig. \ref{fig:SoftIC1} is ignorant of the underlying channel model, its basic building blocks are channel-model-dependent. In particular, interference cancellation requires the contribution of the interference to be additive, i.e., a linear model channel as in \eqref{eqn:Linear channel}, as well as full \ac{csi}, in order to cancel the contribution of the interference. Soft detection requires complete knowledge of the channel input-output relationship in order to estimate the conditional probabilities via \eqref{eqn:CondDist2}. 

\begin{figure}
		\centering
		\includegraphics[width =\figWidth]{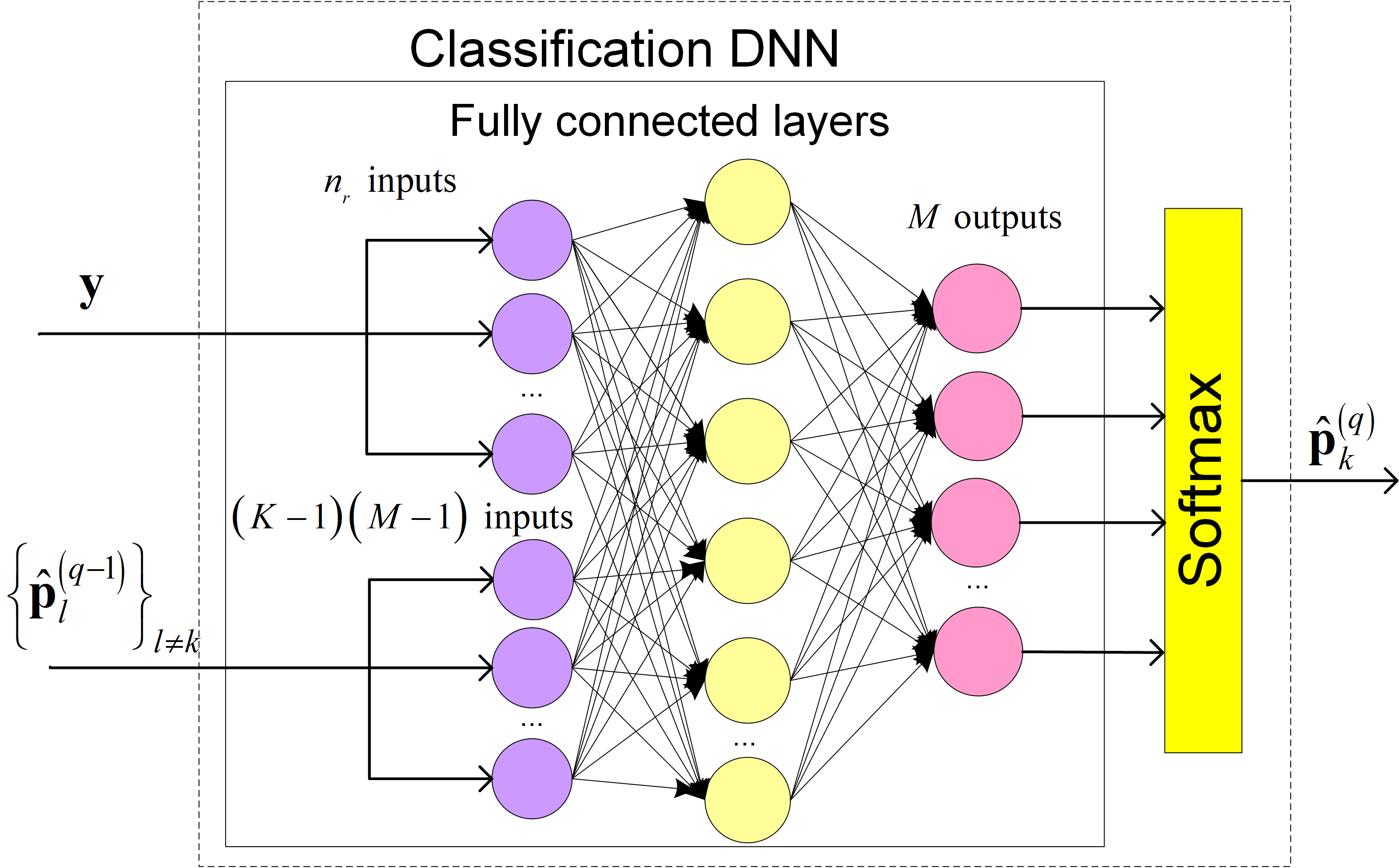} 
		\vspace{-0.2cm}
		\caption{Conditional probability estimation network model. Here, the network has $\CnstSize$ output nodes, representing classification with $\CnstSize$ possible classes.}
		\label{fig:DNN1}
	\end{figure}

\begin{figure}
		\centering
		\includegraphics[width = \columnwidth]{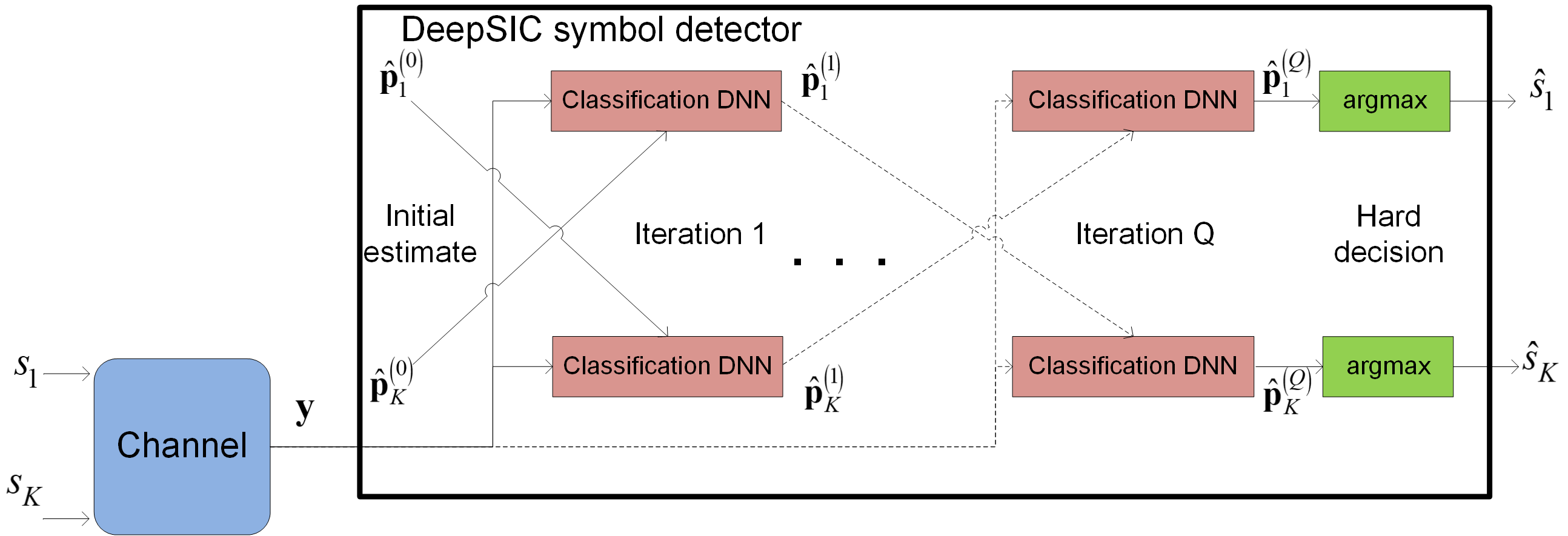} 
		\vspace{-0.2cm}
		\caption{Deep\ac{sic} illustration.}
		\label{fig:DeepSoftIC1}
	\end{figure}

Although each of these basic building blocks consists of two sequential procedures which are completely channel-model-based, we note that the purpose of these computations is to carry out a classification task. In particular, the $k$th building block of the $q$th iteration, $k \in \NusersSet$, $q  \in \NiterSet$, produces $\hat{\myVec{p}}_k^{(q)}$, which is an estimate of the conditional distribution of $S_k[i]$ given $\myY[i] = \myVec{y}$ based on $\{\hat{\myVec{p}}_l^{(q-1)}\}_{l\neq k}$. Such computations are naturally implemented by classification \acp{dnn}, e.g., fully-connected networks with softmax output layer. 
\label{txt:Classifier}
An illustration of such a network implementing the $k$th basic block of the $q$th iteration is depicted in Fig. \ref{fig:DNN1}. 
\textcolor{NewColor}{
Specifically, in Fig. \ref{fig:DNN1} we depict a fully-connected multi-layered network with $\CnstSize$ output nodes and a softmax output layer, whose inputs are the $\Nantennas \times 1$ channel output $\myVec{y}$ and the previous interference conditional distribution estimates $\{\hat{\myVec{p}}_{l}^{(q-1)}\}_{l\neq k}$. While the latter consists of $(\Nusers -1) \CnstSize$ entries, it can be represented using $(\Nusers-1)(\CnstSize-1)$ values, since the sum of the entries of each such probability vector is one. When trained to minimize the cross-entropy loss, the building block \ac{dnn} implements a neural classifier for  $\CnstSize$ possible labels, i.e., constellation points.  As such, the entries of the output vector of the softmax layer represents an estimate of the conditional probability for each possible symbol conditioned on $\myVec{y}$ and $\{\hat{\myVec{p}}_{l}^{(q-1)}\}_{l\neq k}$.} 
Embedding these \acl{ml} based conditional distribution computations into the iterative \ac{sic} block diagram in Fig. \ref{fig:SoftIC1} yields the overall receiver architecture depicted in Fig. \ref{fig:DeepSoftIC1}. We set the initial estimates $\{\hat{\myVec{p}}_k^{(0)}\}_{k=1}^{\Nusers}$ to represent a uniform distribution, i.e., $\big(\hat{\myVec{p}}_k^{(0)}\big)_m = \frac{1}{\CnstSize}$ for each $m \in \{1,2,\ldots,\CnstSize\}$ and $k \in \NusersSet$. We leave the study of different initial estimates and their effect on the overall receiver performance for future research. The resulting data-driven implementation of Algorithm \ref{alg:Algo1} is repeated below as Algorithm \ref{alg:AlgoDeepSIC}. Note that the model-based Steps \ref{stp:MF1}-\ref{stp:SoftDec} of Algorithm \ref{alg:Algo1} whose purpose is to estimate the conditional distributions, are replaced with the \acl{ml} based conditional distribution estimation Step \ref{stp:DeepSoftDec} in Algorithm \ref{alg:AlgoDeepSIC}.

\begin{algorithm}
	\caption{Deep Soft Interference Cancellation (Deep\ac{sic})}
	\label{alg:AlgoDeepSIC}
	\begin{algorithmic}[1]
		\STATE \underline{Input}: Channel output $\myVec{y}$.
		\STATE \underline{Initialization}: Set $q=1$, and generate an initial guess of the conditional distributions $\{\hat{\myVec{p}}_k^{(0)}\}_{k =1}^{\Nusers}$.  
		\STATE \label{stp:DeepSoftDec} {\em Conditional distribution estimation:} For each $k \in \NusersSet$, estimate the conditional distribution $\hat{\myVec{p}}_k^{(q)}$ from  $\myVec{y}$ and $\{\hat{\myVec{p}}_k^{(q-1)}\}_{l\neq k}$ using the $(q,k)$th classification \ac{dnn}.
		\STATE Set $q := q+1$. If $q \le \Niter $ go to Step \ref{stp:DeepSoftDec}.	
		
		\STATE  \underline{Output}: Hard detected output $\hat{\myVec{s}}$, obtained via \eqref{eqn:HardDet}.
	\end{algorithmic}
\end{algorithm} 

A major advantage of using classification \acp{dnn} as the basic building blocks in Fig. \ref{fig:DeepSoftIC1} stems from the fact that such \acl{ml} methods are capable of accurately computing conditional distributions in   complex non-linear setups without requiring a-priori knowledge of the channel model and its parameters. Consequently, when these building blocks are trained to properly implement their classification task, the  receiver essentially realizes iterative soft interference cancellation for arbitrary channel models in a data-driven fashion. In the following subsection we discuss how to train these classification \acp{dnn}.

\subsection{Training the \acp{dnn}}
\label{subsec:Training}
\vspace{-0.1cm}	
In order for the \acl{ml} based receiver structure of Fig. \ref{fig:DeepSoftIC1} to  reliably implement joint detection, its building block classification \acp{dnn} must be properly trained. Here, we consider two possible approaches to train the receiver based on the training set of $\Ntraining$ pairs of channel inputs and their corresponding outputs $\{\tilde{\myVec{s}}_j, \tilde{\myVec{y}}_j \}_{j=1}^{\Ntraining}$: {\em End-to-end training}, and {\em sequential training}.

\label{txt:endtoend}
{\bf End-to-end training}: The first approach jointly trains the entire network, i.e., all the building block \acp{dnn}. 
\textcolor{NewColor}{
While the output of each building block is an $\CnstSize\times 1$ vector, as illustrated in Fig.~\ref{fig:DNN1}, the output of the overall interconnection of these \acp{dnn} is the set of conditional distributions  $\{\hat{\myVec{p}}_k^{(\Niter)}\}_{k=1}^{\Nusers}$ as illustrated in Fig. \ref{fig:DeepSoftIC1}. Since each vector $\hat{\myVec{p}}_k^{(\Niter)}$ is used to estimate $S_k[i]$, we use the sum-cross entropy loss as the training objective.} 
 Let 
$\myVec{\theta}$ represent the parameters of the entire network, and 
$\hat{\myVec{p}}_k^{(\Niter)}(\myVec{y}, \alpha; \myVec{\theta} )$ be the entry of $\hat{\myVec{p}}_k^{(\Niter)}$ corresponding to $S_k[i] = \alpha$ when the input to the network is $\myVec{y}$ and its parameters are $\myVec{\theta}$.
The sum-cross entropy loss over the training set $\{\tilde{\myVec{s}}_j,\tilde{\myVec{y}}_j\}_{j=1}^{\Ntraining}$ can be written as
\begin{equation}
\label{eqn:SumCE}
\mathcal{L}_{\rm SumCE} (\myVec{\theta})= \frac{1}{\Ntraining}\sum_{j=1}^{\Ntraining}
\sum_{k=1}^{\Nusers} -\log \hat{\myVec{p}}_k^{(\Niter)}\big(\tilde{\myVec{y}}_j, (\tilde{\myVec{s}}_j)_k ; \myVec{\theta}\big).
\end{equation} 

Training the receiver in Fig. \ref{fig:DeepSoftIC1} in this end-to-end manner based on the objective \eqref{eqn:SumCE} jointly updates the coefficients of all the $\Nusers \cdot \Niter$ building block \acp{dnn}. Since for a large number of users, training so many parameters simultaneously is expected to require a large number of input-output pairs, we further propose a sequential training approach detailed next. 

{\bf Sequential training}: To allow the network to be trained with a reduced number of training samples, we note that the goal of each building block \ac{dnn} does not depend on the iteration index: The $k$th building block of the $q$th iteration outputs a soft estimate of $S_k[i]$ for each $q \in \NiterSet$; this estimation is iteratively refined as the iteration index grows. Therefore, each building block \ac{dnn} can be trained individually, by minimizing the conventional cross entropy loss. To formulate this objective, let  
$\myVec{\theta}_{k}^{(q)}$ represent the parameters of the $k$th \ac{dnn} at iteration $q$, and write 
$\hat{\myVec{p}}_k^{(q)}\big(\myVec{y}, \{\hat{\myVec{p}}_l^{(q-1)}\}_{l\neq k}, \alpha; \myVec{\theta}_{k}^{(q)}\big)$ as the entry of $\hat{\myVec{p}}_k^{(q)}$ corresponding to $S_k[i] = \alpha$ when the \ac{dnn} parameters are $\myVec{\theta}_{k}^{(q)}$ and its inputs are $\myVec{y}$ and  $\{\hat{\myVec{p}}_l^{(q-1)}\}_{l\neq k}$. The cross entropy loss is given by
 \begin{equation}
 \label{eqn:CE}
 \mathcal{L}_{\rm CE}(\myVec{\theta}_{k}^{(q)})  = \frac{1}{\Ntraining}\sum_{j=1}^{\Ntraining}
  -\log \hat{\myVec{p}}_k^{(q)}\big(\tilde{\myVec{y}}_j, \{\hat{\myVec{p}}_{j,l}^{(q-1)}\}_{l\neq k}, (\tilde{\myVec{s}}_j)_k ; \myVec{\theta}_{k}^{(q)}\big),
 \end{equation}	
 where $\{\hat{\myVec{p}}_{j,l}^{(q-1)}\}$ represent the estimated probabilities associated with $\tilde{\myVec{y}}_j$ computed at the previous iteration.
 The problem with training each \ac{dnn} individually to minimize \eqref{eqn:CE} is that the soft estimates $\{\hat{\myVec{p}}_{j,l}^{(q-1)}\}$ are not provided as part of the training set. This challenge can be tackled by training the \acp{dnn} corresponding to each layer in a sequential manner, where for each layer the outputs of the trained \ac{dnn} corresponding to the previous iterations are used to generate the soft estimates fed as training samples. This process is summarized below as Algorithm \ref{alg:Algo2}.
 
Sequential training uses the $\Ntraining$ input-output pairs to train each \ac{dnn} individually. Compared to the end-to-end training that utilizes the training samples to learn the complete set of parameters, which can be quite large, sequential training uses the same set of input-output pairs to learn a significantly smaller number of parameters, reduced by a factor of $\Nusers\cdot \Niter$, multiple times. Consequently, this approach is expected to require a much smaller number of training samples, at the cost of a longer learning procedure for a given training set, due to its sequential operation, and possible performance degradation as the building blocks are not jointly trained. This behavior is numerically demonstrated in the simulation study detailed in Section \ref{sec:Sims}, where the performance gap between sequential training and end-to-end training is shown to be relatively minor.
	
\begin{algorithm}
	\caption{Sequential Training Algorithm}
	\label{alg:Algo2}
	\begin{algorithmic}[1]
		\STATE \underline{Input}: Training samples$\{\tilde{\myVec{s}}_j,\tilde{\myVec{y}}_j\}_{j=1}^{\Ntraining}$.
		\STATE \underline{Initialization}: Set $q=1$, generate an initial guess of the conditional distributions $\{\hat{\myVec{p}}_k^{(0)}\}_{k =1}^{\Nusers}$, and set $\hat{\myVec{p}}_{j,k}^{(0)} = \hat{\myVec{p}}_k^{(0)}$ for each $k \in \NusersSet$ and $j \in \{1,2,\ldots,\Ntraining\}$. 
		\FOR{each $k \in \NusersSet$ \label{stp:InitLayer} } 
		\STATE   Randomize initial weights $\myVec{\theta}_k^{(q)}$.
		\STATE \label{stp:Learn}  Train $\myVec{\theta}_k^{(q)}$ to minimize \eqref{eqn:CE}.
		\STATE \label{stp:GenTrain}  Feed $\big\{\tilde{\myVec{y}}_j,\{\hat{\myVec{p}}_{j,l}^{(q-1)}\}_{l\neq k}\big\}_{j=1}^{\Ntraining}$ to the trained \ac{dnn}, producing  $\{\hat{\myVec{p}}_{j,k}^{(q)}\}_{j=1}^{\Ntraining}$.
		\ENDFOR
		\STATE Set $q := q+1$. If $q \le \Niter $ go to Step \ref{stp:InitLayer}.	
		\STATE  \underline{Output}: Trained network parameters $\myVec{\theta} = \{\myVec{\theta}_k^{(q)}\}$.
	\end{algorithmic}
\end{algorithm}

\vspace{-0.2cm}
\subsection{Discussion}
\label{subsec:Discussion}
\vspace{-0.1cm}
Deep\ac{sic} learns to implement the iterative \ac{sic} algorithm from training. Consequently, once trained, it shares the main advantages of the model-based algorithm: For scenarios in which iterative \ac{sic} is applicable, i.e., linear channels of the form \eqref{eqn:Linear channel}, the performance of deep\ac{sic} is expected to approach that of the optimal \ac{map} detector. This behavior is numerically observed in the simulation study detailed in Section \ref{sec:Sims}.

\textcolor{NewColor}{
Similarly to the model-based iterative \ac{sic} algorithm from which DeepSIC originates, the computational complexity of applying Deep\ac{sic} grows linearly with the number of users. This makes DeepSIC applicable in \ac{mimo} scenarios where conventional \ac{map} detection is infeasible. Furthermore, as DeepSIC consists of an interconnection of relatively compact \acp{dnn}, it shares the advantage of \ac{dnn}-based receivers over iterative model-based receivers  in terms of inference speed \cite{Balatsoukas:19}. Explicitly characterizing the complexity of DeepSIC is challenging due to the inherent difficulty in quantifying the complexity of training and applying \acp{dnn}, which is heavily dependent on the number of parameters and the learning algorithm. 
In particular, by letting $\Nparameters$ be the number of parameters in each building block \ac{dnn}, the overall number of parameters of Deep\ac{sic} is $\Nparameters \cdot\Nusers\cdot \Niter$.  Additionally, using sequential training, the training set is required to adapt only $\Nparameters$ parameters, as each building block is trained individually. This indicates that Deep\ac{sic} can learn to implement iterative \ac{sic} using a relatively small number of training samples, allowing it to achieve improved performance over previously proposed deep receivers with smaller training sets, as we numerically demonstrate in Subsection \ref{subsec:Linear}. Additionally, the ability of DeepSIC to accurately train with small data sets  facilitates tracking of dynamic channel conditions via self-supervised online training by exploiting the inherent structure of communication protocols \cite[Sec. IV]{Shlezinger:19a}, as we demonstrate in Subsection \ref{subsec:Online}.}


In addition to its ability to implement iterative \ac{sic} without prior knowledge of the channel model and its parameters, Deep\ac{sic} has two main advantages over the model-based algorithm from which it originates: 
First, since Deep\ac{sic} learns to cancel the interference from training, and does not assume that its contribution is additive, it is applicable also in non-linear channels. Iterative \ac{sic}, which attempts to cancel the interference by subtracting its estimate from the channel output, results in increased errors in  non-linear channels. 
Furthermore, even in linear scenarios where iterative \ac{sic} is applicable, its performance is highly sensitive to inaccurate \ac{csi}. By exploiting the known generalization properties of \acp{dnn}, Deep\ac{sic} is capable of operating reliably when trained under different channel parameters, which is equivalent to having inaccurate \ac{csi}. These advantages are clearly observed in our numerical study in Section~ \ref{sec:Sims}.

To identify the advantages of Deep\ac{sic} over previously proposed \acl{ml} based detectors, we first recall that, as discussed in the introduction, these data-driven receivers can be divided into two main types: 
The first family of deep receivers implements symbol detection using a single conventional network, generally treated as a black box. While their architecture  can account for some a-priori knowledge of the scenario, such as \ac{ofdm} signaling \cite{Ye:18,Mosleh:18}, channel memory \cite{Farsad:18,Liao:19}, and the presence of low resolution quantizers \cite{Balevi:19,Shlezinger:19c}, the design is typically not based on established detection algorithms.  Compared to such deep receivers, Deep\ac{sic}, which learns to implement only the model-based computations of an established algorithm, has fewer parameters, and can thus be trained using smaller training sets, allowing it to be quickly retrained in the presence of dynamic environments. Furthermore, unlike black box \acp{dnn}, the architecture of Deep\ac{sic} is interpretable, and, when properly trained, it is expected to achieve the \ac{map}-comparable performance of iterative \ac{sic}.     

The second family of \acl{ml}-driven \ac{mimo} receivers unfolds a model-based iterative optimization algorithm for finding the \ac{map} solution \eqref{eqn:MAP}, such as projected gradient descent,  into  a \ac{dnn}, see, e.g., \cite{Wiesel:19,Corlay:18,He:19a,Takabe:19} as well as \cite[Sec. II]{Balatsoukas:19}. However, these methods assume a linear channel model of the form \eqref{eqn:Linear channel}. Furthermore, these previous receivers typically require \ac{csi}, obtained either from a-priori knowledge or via channel estimation as in \cite{He:19b}. Unlike these previous receivers, Deep\ac{sic}, which is also based on a model-based algorithm, is independent of the channel model, and can efficiently learn to detect in a wide variety of channel conditions, ranging from linear Gaussian channels to non-linear Poisson channels, as numerically demonstrated in Section \ref{sec:Sims}.

Finally, we note that the architecture of Deep\ac{sic}, which is depicted in Fig. \ref{fig:DeepSoftIC1}, is related to the concepts of deep mutual learning \cite{Zhang:18}, as well as plug-and-play methods \cite{Venkatakrishnan:13}. In particular, in deep mutual learning, a set of relatively small \acp{dnn} are each trained individually while allowing to share some information between intermediate layers to facilitate training. By considering the $\Niter$ building blocks corresponding to each user as a single \ac{dnn}, Deep\ac{sic} can be considered as a form of mutual learning between those networks. However, while conventional mutual learning is based on heuristic arguments, Deep\ac{sic} arises from an established detection algorithm, which is particularly suitable for the problem of \ac{mimo} detection. As such, the building blocks of Deep\ac{sic} are not arbitrary layers, but classification networks designed and trained to implement the corresponding computation of iterative \ac{sic} in a data-driven fashion. Plug-and-play methods implement regularized optimization algorithms involving proximal operations by replacing these computations with denoiser \acp{dnn}, allowing the optimization process to be carried out in a data-driven manner without analytically accounting for the regularization. Our approach in designing Deep\ac{sic} thus bears some similarity to plug-and-play methods, in the sense that both approaches integrate \acp{dnn} into an established algorithm by replacing some specific model-based computations with \acp{dnn}. While their underlying rationale may be similar, Deep\ac{sic} and plug-and-play methods are fundamentally different in the algorithm from which they arise, as well as their target application, i.e., \ac{mimo} detection versus inverse problems in image processing.

	\vspace{-0.2cm}
	\section{Numerical Study}
	\label{sec:Sims}
	\vspace{-0.1cm}
	In the following section we numerically evaluate Deep\ac{sic} in several relevant multiuser \ac{mimo} detection scenarios. We first consider linear Gaussian channels in Subsection \ref{subsec:Linear}, for which conventional model-based iterative \ac{sic} as well as the majority of previously proposed \ac{dnn}-based \ac{mimo} detectors are applicable. Then, in Subsection \ref{subsec:NonLinear}, we  demonstrate the performance gains of Deep\ac{sic} in two common non-linear scenarios: Quantized Gaussian setups and Poisson channels. Next, in Subsection \ref{subsec:TrainSim} we compare the methods for training Deep\ac{sic} discussed in Subsection \ref{subsec:Training}, \textcolor{NewColor}{and in Subsection \ref{subsec:Online} we evaluate DeepSIC in block-fading channels.} 
	
\begin{figure}
		\centering
		\includegraphics[width = 0.8\columnwidth]{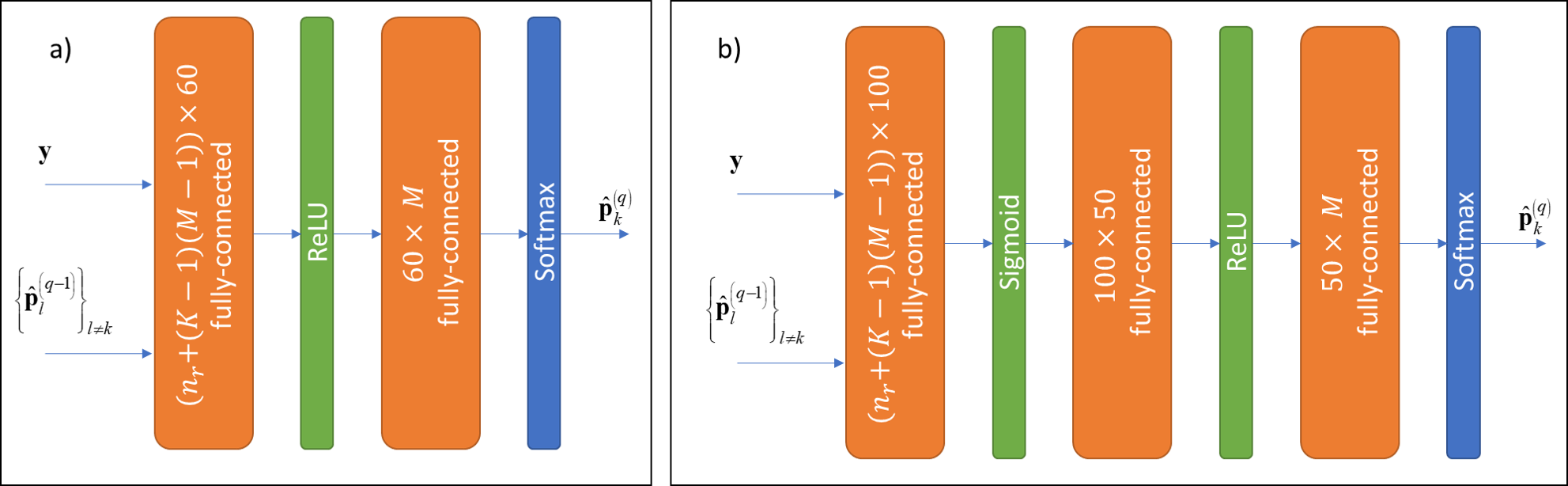} 
		\vspace{-0.2cm}
		\caption{\ac{dnn} architectures used as DeepSIC building blocks for: $a)$ end-to-end training; $b)$ sequential training.}
		\label{fig:SimArch1}
	\end{figure}

	Unless stated otherwise, we trained  Deep\ac{sic}  using the ADAM optimizer with a relatively small training set of $5000$ training samples, and tested over $20000$ symbols.	
	\textcolor{NewColor}{
	The motivation for using small training sets is to demonstrate the ability of Deep\ac{sic} to train with a sample set of the order of a preamble sequence, e.g., \cite[Ch. 17]{Dahlman:10}, indicating its feasibility to exploit the structure induced by communication protocols to adapt in dynamic environments, as we demonstrate in Subsection \ref{subsec:Online}. 
	We simulate DeepSIC with both end-to-end training as well as sequential training. Since the latter strategy sequentially adapts subsets of the building blocks, it can  tune a larger number of parameters using the same training set compared to end-to-end training,  where all the building blocks are jointly trained. 			
	Consequently, in the implementation of the \ac{dnn}-based building blocks of Deep\ac{sic} depicted in Fig. \ref{fig:DNN1}, we used a different fully-connected network for each training method. In particular, for end-to-end training we used a compact network consisting of a  $(\Nantennas + (\Nusers - 1)(\CnstSize-1)) \times 60$ layer followed by ReLU activation and a $60 \times \CnstSize$  layer, as illustrated in Fig. \ref{fig:SimArch1}(a). 
	For sequential training, we used three fully-connected layers:  An $(\Nantennas + (\Nusers - 1)(\CnstSize-1)) \times 100$ first layer, a $100 \times 50$ second layer, and a $50 \times \CnstSize$ third layer, where a sigmoid and a ReLU intermediate activation functions were used, respectively.  The resulting \ac{dnn} is illustrated in Fig. \ref{fig:SimArch1}(b).  
	}
	We note that the different training methods are also compared with the same \ac{dnn} structures in Subsection \ref{subsec:TrainSim}, allowing to determine which method should be used based on the training set size.

	\vspace{-0.2cm}
	\subsection{Linear Gaussian Channels}
	\label{subsec:Linear}
	\vspace{-0.1cm}
	We first consider a linear \ac{awgn} channel whose input-output relationship is given by \eqref{eqn:Linear channel}. Recall that the model-based algorithm from which Deep\ac{sic} originates, i.e., iterative \ac{sic}, as well as previously proposed unfolding-based data-driven \ac{mimo} receivers \cite[Sec. II]{Balatsoukas:19}, are all designed for such channels. Consequently, the motivation of the following study is to compare Deep\ac{sic} in terms of performance and robustness to competing detectors in a scenario for which these previous schemes are applicable. 
	In particular, we evaluate the \ac{ser} of the following \ac{mimo} detectors:
	\begin{itemize}
		\item The \ac{map} detector, given by \eqref{eqn:MAP}.
		\item The iterative \ac{sic} algorithm detailed in Algorithm \ref{alg:Algo1}.
		\item Deep\ac{sic} with the sequential training method, referred to in the following as {\em Seq. Deep\ac{sic}}.
		\item Deep\ac{sic} with end-to-end training based on the sum cross entropy loss \eqref{eqn:SumCE}, referred to henceforth as {\em E2E Deep\ac{sic}}.
		\item The unfolding-based {\em DetNet} \ac{mimo} detector proposed in \cite{Wiesel:19}.
	\end{itemize}
	
	The model-based \ac{map} and iterative \ac{sic} detectors, as well as DetNet \cite{Wiesel:19}, all require \ac{csi}, and specifically, accurate knowledge of the channel matrix $\myMat{H}$.  Deep\ac{sic} operates without a-priori knowledge of the channel model and its parameters, learning the detection mapping from a training set sampled from the considered input-output relationship. In order to compare the robustness of the  detectors to \ac{csi} uncertainty, we also evaluate them when the receiver has access to an estimate of $\myMat{H}$ with entries corrupted by i.i.d. additive Gaussian noise whose variance is given by $\SigE$ times the magnitude of the corresponding entry, where $\SigE > 0$ is referred to as the {\em error variance}. For Deep\ac{sic}, which is model-invariant, we compute the \ac{ser} under \ac{csi} uncertainty by using a training set whose samples are randomized from a channel in which the true $\myMat{H}$ is replaced with its noisy version. 
	
\begin{figure}
		\centering
		\includegraphics[ width=\figWidth]{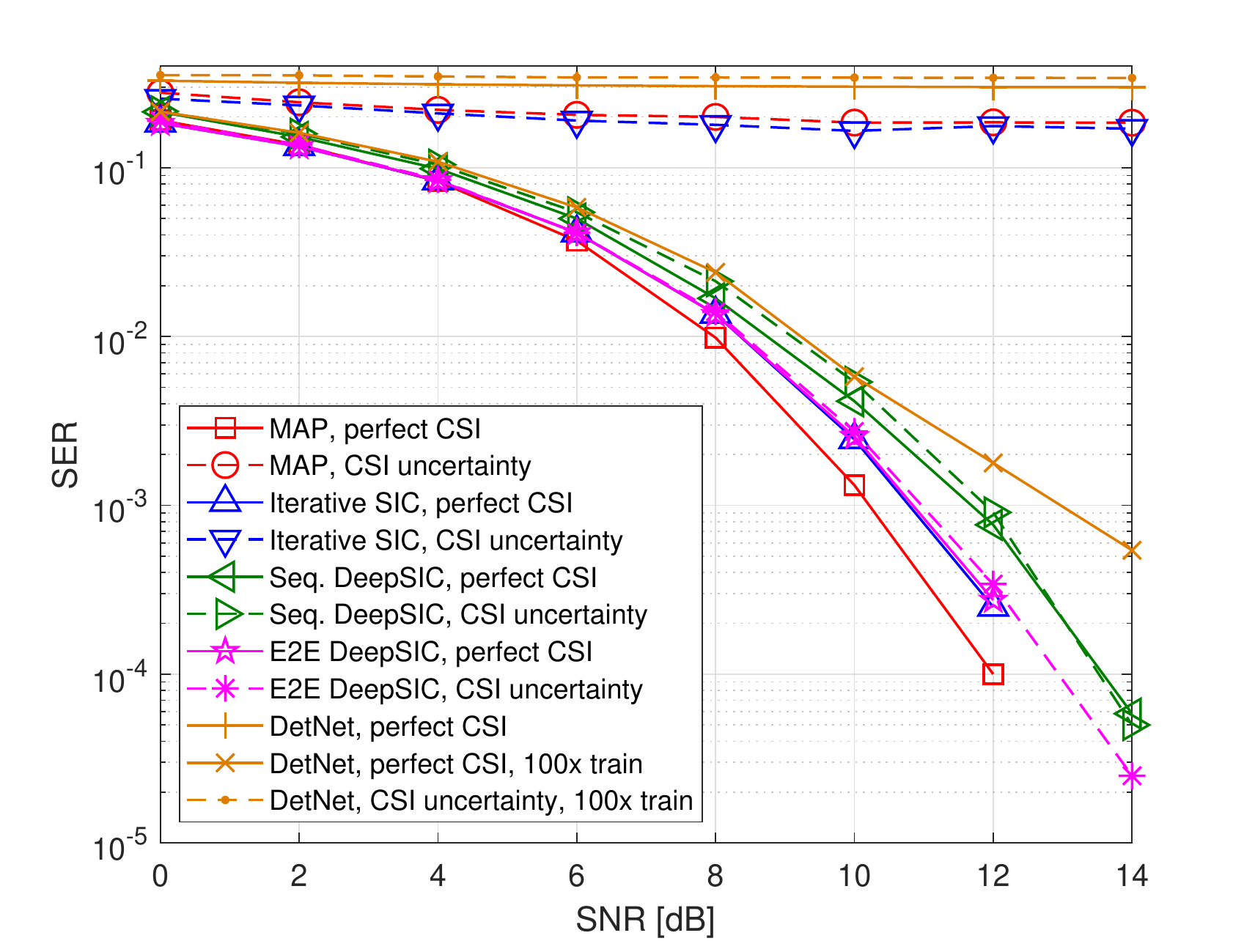}
		\figSpace
		\caption{\textcolor{NewColor}{\ac{ser} versus \ac{snr} of DeepSIC compared to the model-based iterative \ac{sic} and the data-driven DetNet of \cite{Wiesel:19}, $6\times6$ linear channel with \ac{awgn}. For DeepSIC and DetNet, {\em Perfect CSI} implies that the receiver is trained and tested using samples from the same channel, while under {\em CSI uncertainty} they are trained using samples from a set of different channels.}
		}
		\label{fig:AWGN6} 
	\end{figure}

\begin{figure}
		\centering
		\includegraphics[ width=\figWidth]{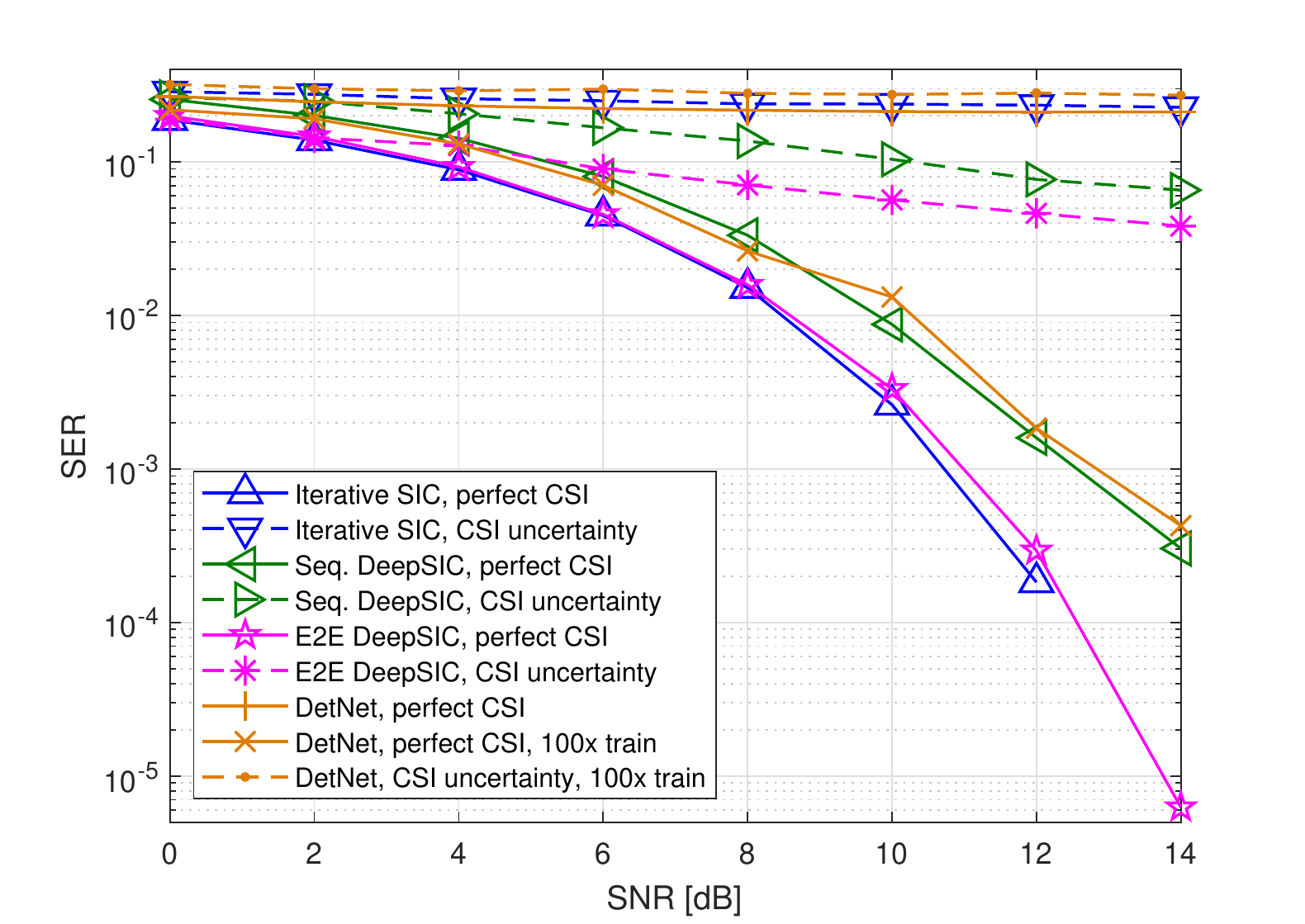}
		\figSpace
		\caption{\textcolor{NewColor}{\ac{ser} versus \ac{snr} of DeepSIC compared to the model-based iterative \ac{sic} and the data-driven DetNet of \cite{Wiesel:19}, $32\times32$ linear channel with \ac{awgn}. For DeepSIC and DetNet, {\em Perfect CSI} implies that the receiver is trained and tested using samples from the same channel, while under {\em CSI uncertainty} they are trained using samples from a set of different channels.}
		}
		\label{fig:AWGN32} 
	\end{figure}

	We simulate two linear Gaussian channels: A $6 \times 6$ channel, i.e., $\Nusers = 6$ users and $\Nantennas = 6$ receive antennas, and a $32 \times 32$ setup. Since the computational complexity of the \ac{map} rule grows exponentially with $\Nusers$, it is  simulated only for the $6\times 6$ channel. 
	\textcolor{NewColor}{
	Consequently, simulating the $6 \times 6$ channel allows us to compare the error rate achieved by DeepSIC to that of the optimal \ac{map} rule, while the purpose of the $32 \times 32$ setup is demonstrate the feasibility of DeepSIC in large multi-user \ac{mimo} systems, where the \ac{map} rule become computationally prohibitive.} 
	The symbols are randomized from a \ac{bpsk} constellation, namely, $\mySet{S} = \{-1, 1\}$ and $\CnstSize = |\mySet{S}| = 2$. The channel matrix $\myMat{H}$ models spatial exponential decay, and its entries are given by
	\begin{equation}
	\left( \myMat{H}\right)_{i,j} = e^{-|i-j|}. \qquad i \in \{1,\ldots, \Nantennas\}, j \in \{1,\ldots, \Nusers\}. 
	\label{eqn:ChannelMat}
	\end{equation}
	For each channel, the \ac{ser} of the considered receivers is evaluated for both perfect \ac{csi}, i.e., $\SigE = 0$, as well as \ac{csi} uncertainty, for which we use $\SigE = 0.1$ and $\SigE = 0.75$ for the $6 \times 6$ and the $32\times 32$ channels, respectively.  The numerically evaluated \ac{ser} values versus the \ac{snr}, defined as $1/\SigW$, are depicted in Figs. \ref{fig:AWGN6}-\ref{fig:AWGN32} for the $6\times 6$ case and the $32\times 32$ channel, respectively. 
	
	Observing Fig. \ref{fig:AWGN6}, we note that the performance of Deep\ac{sic} with end-to-end training approaches that of the model-based iterative \ac{sic} algorithm, which is within a small gap of the optimal \ac{map} performance. This demonstrates the ability of Deep\ac{sic} to implement iterative \ac{sic} in a data-driven fashion. The sequential training method, whose purpose is to allow Deep\ac{sic} to train with smaller data sets compared to end-to-end training, also achieves \ac{ser} which is comparable to iterative \ac{sic}. In particular, the optimal \ac{map} detector with perfect \ac{csi} achieves \ac{ser} of $10^{-3}$ at \ac{snr} of approximately $10$ dB, while iterative \ac{sic} and E2E Deep\ac{sic} require $11$ dB \ac{snr} and Seq. Deep\ac{sic} needs an \ac{snr} of at least $12$ dB to achieve the same \ac{ser} value. The fact that E2E Deep\ac{sic} approaches the performance of iterative \ac{sic} with perfect \ac{csi} is also observed for the $32\times32$ channel in Fig. \ref{fig:AWGN32}, indicating the applicability of Deep\ac{sic} in relatively large multiuser \ac{mimo} networks, where \ac{map}-based joint detection is computationally infeasible.  We note that the \ac{snr} gap of sequential training compared to end-to-end training is more dominant in the $32\times32$, and is approximately $1.5$ dB for \ac{ser} of $10^{-3}$. The ability of Deep\ac{sic} to learn its mapping from a small training set becomes notable when comparing its performance to DetNet: For the same training set of size $\Ntraining = 5000$, DetNet fails to properly adapt and achieves very poor \ac{ser} performance. Only when provided a hundred times more training samples, denoted {\em 100x train} in Figs.\ref{fig:AWGN6}-\ref{fig:AWGN32}, DetNet achieves \ac{ser} values within a small gap of that achieved by Seq. Deep\ac{sic} with $1\%$ of the training set.
	
	Figs. \ref{fig:AWGN6}-\ref{fig:AWGN32} indicate that the performance Deep\ac{sic} is comparable to iterative \ac{sic} with accurate \ac{csi}. However, in the presence of \ac{csi} uncertainty, Deep\ac{sic} is observed to substantially outperform the model-based iterative \ac{sic} and \ac{map} receivers, as well as DetNet operating with a noisy version of $\myMat{H}$ and trained with a hundred times more inaccurate training samples. In particular, it follows from Fig. \ref{fig:AWGN6} that a relatively minor estimation error of variance   $\SigE = 0.1$ severely deteriorates the performance of the model-based methods, while the proposed data-driven Deep\ac{sic} is hardly affected by the same level of \ac{csi} uncertainty. In Fig. \ref{fig:AWGN32}, in which the relatively large error variance $\SigE = 0.75$ is used, we observe that while iterative \ac{sic} is inapplicable, Deep\ac{sic} is still capable of achieving \ac{ser} values which decrease below $10^{-1}$ for \acp{snr} larger than $10$ dB in the presence of notable \ac{csi} errors. However, the fact that it is trained here using a training set which is not only relatively small, but also quite noisy, induces a notable loss compared to accurate training. Nonetheless, Deep\ac{sic} is still shown to be far more robust to \ac{csi} uncertainty compared to the model-based iterative \ac{sic}.

%
	
	\vspace{-0.2cm}
	\subsection{Non-Linear Channels}
	\label{subsec:NonLinear}
	\vspace{-0.1cm}	
	Here, we evaluate Deep\ac{sic} in communication channels which are not modeled as a linear system whose output is corrupted by \ac{awgn}. The purpose of this study is to demonstrate that, even though the model-based iterative \ac{sic} algorithm is derived assuming a memoryless linear \ac{awgn} channel, its data-driven adaptation can be reliably applied in a much broader family of relevant scenarios. Since the unfolding-based DetNet receiver of \cite{Wiesel:19} is designed for linear \ac{awgn} setups, here we compare Deep\ac{sic} only to the model-based \ac{map} and the iterative \ac{sic} receivers. 
	
	We begin with a quantized Gaussian channel, which typically models wireless communications in the presence of low-resolution quantizers \cite{Shlezinger:19b}. The channel output undergoes a $2$ bits uniform quantization mapping over the support $[-4, 4]$, given by
	\begin{equation*}
	q(y) = \begin{cases}
	{\rm sign}(y)   & |y| < 2 \\
		3 \cdot {\rm sign}(y)    & |y| > 2.
	\end{cases}
	\end{equation*}  	
	Consequently, using the notations in \eqref{eqn:Linear channel}, the channel input-output relationship can be written as
	\begin{equation}
	\label{eqn:QGaussian}
	\myY[i] = q\left( \myMat{H} \myS[i] + \myVec{W}[i]\right), 
	\end{equation}
	where the quantization in \eqref{eqn:QGaussian} is carried out entry-wise. 
	In particular, we consider a receiver with $\Nantennas = 4$ antennas serving $\Nusers = 4$ users, each transmitting i.i.d. \ac{bpsk} symbols. The entries of the channel matrix $ \myMat{H}$ are given by \eqref{eqn:ChannelMat}. 
	
\begin{figure}
	\centering
	\includegraphics[ width=\figWidth]{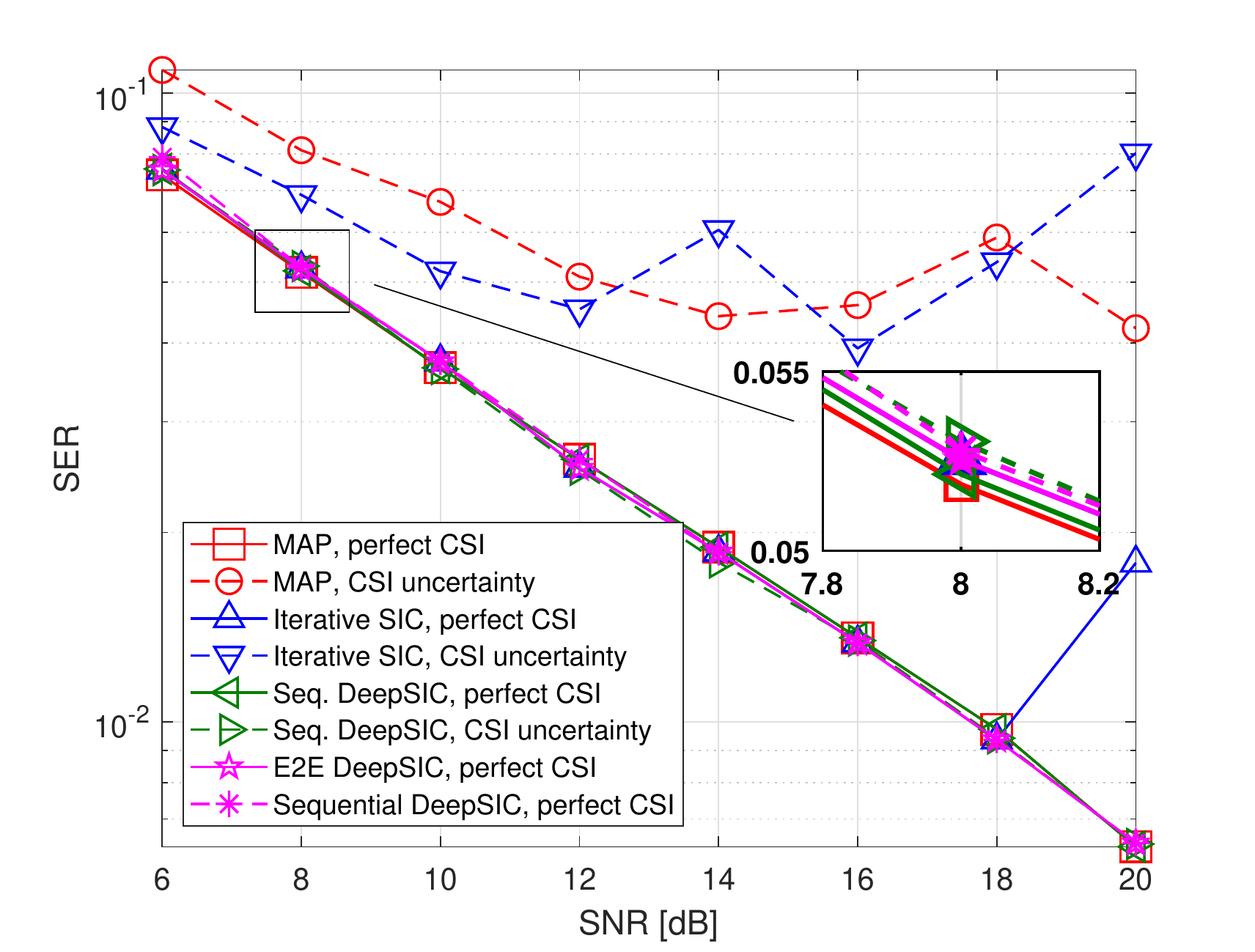}
	\figSpace
	\caption{\textcolor{NewColor}{\ac{ser} versus \ac{snr} of DeepSIC compared to the model-based \ac{map} rule and iterative \ac{sic} method, 
			$4\times4$ quantized Gaussian channel. For DeepSIC, {\em Perfect CSI} implies that the receiver is trained and tested using samples from the same channel, while under {\em CSI uncertainty} it is trained using samples from a set of different channels.}
	}
	\label{fig:Qgaussian4} 
	\end{figure}

	In Fig. \ref{fig:Qgaussian4} we compare the  \ac{ser} achieved by Deep\ac{sic} with both end-to-end and sequential training to the performance of the model-based \ac{map} and iterative \ac{sic} receivers versus ${\rm SNR} \in [6,20]$ dB. As in the previous subsection, we consider both the cases in which the receiver has perfect \ac{csi} as well as \ac{csi} uncertainty: Under perfect \ac{csi}, the model-based \ac{map} and iterative \ac{sic} detectors have accurate knowledge of $\myMat{H}$ and $\SigW$, while Deep\ac{sic} is trained over samples taken from the same channel model under which it is tested; In \ac{csi} uncertainty the model-based receivers have access to a noisy estimate of $\myMat{H}$ with uncertainty variance $\SigE = 0.1$, and Deep\ac{sic} is trained using samples from the corresponding inaccurate channel model.  
	
	Observing Fig. \ref{fig:Qgaussian4}, we note that the performance of Deep\ac{sic} effectively coincides with that of the optimal \ac{map} rule with perfect \ac{csi}, demonstrating its ability to learn to accurately detect in complex environments. For most considered \ac{snr} values, iterative \ac{sic} with perfect \ac{csi} also approaches the \ac{map} \ac{ser} performance, which settles with the observation in \cite{Wang:17} that iterative soft detection based equalization methods, such as iterative \ac{sic}, achieve excellent performance in quantized Gaussian channels. However, obtaining accurate channel estimation in the presence of low resolution quantization is substantially more challenging compared to conventional linear channels \cite{Shlezinger:19b}, thus in practice, the channel estimates are likely to be inaccurate. 
	\label{txt:Zigzag}
	\textcolor{NewColor}{
	In such cases, it is  shown in Fig. \ref{fig:Qgaussian4} that model-based methods are highly sensitive to inaccurate \ac{csi}, and their error does not monotonically decrease with \ac{snr}. This non-monotonic behavior follows since in some scenarios involving quantized observations, the presence of noise, which causes the discrete channel outputs to change their values at some probability, can reduce the errors induced by \ac{csi} uncertainty. Unlike the model-based receivers, Deep\ac{sic} hardly exhibits any degradation under the same level of uncertainty.}  
		
	Next, we consider a Poisson channel, which typically models free-space optical communications \cite{Khalighi:14}. As in the quantized Gaussian case, we use $\Nusers = 4$ and $\Nantennas = 4$. Here, the symbols are randomized from an on-off keying for which $\mySet{S} = \{0,1\}$. The entries of the channel output are related to the input via the conditional distribution
	\begin{equation}
	\label{eqn:PoissonChannel}
	\left( \myY[i]\right)_j  | \myS[i]  \sim \mathbb{P}\left( \frac{1}{\sqrt{\SigW}}\left( \myMat{H} \myS[i]\right)_j  + 1\right), \qquad j \in \{1,\ldots, \Nantennas\},
	\end{equation}
	where $\mathbb{P}(\lambda)$ is the Poisson distribution with parameter $\lambda > 0$, and the entries of $\myMat{H} \in \mySet{R}^{\Nantennas \times \Nusers}$ are given by \eqref{eqn:ChannelMat}. 
	
\begin{figure}
	\centering
	\includegraphics[ width=\figWidth]{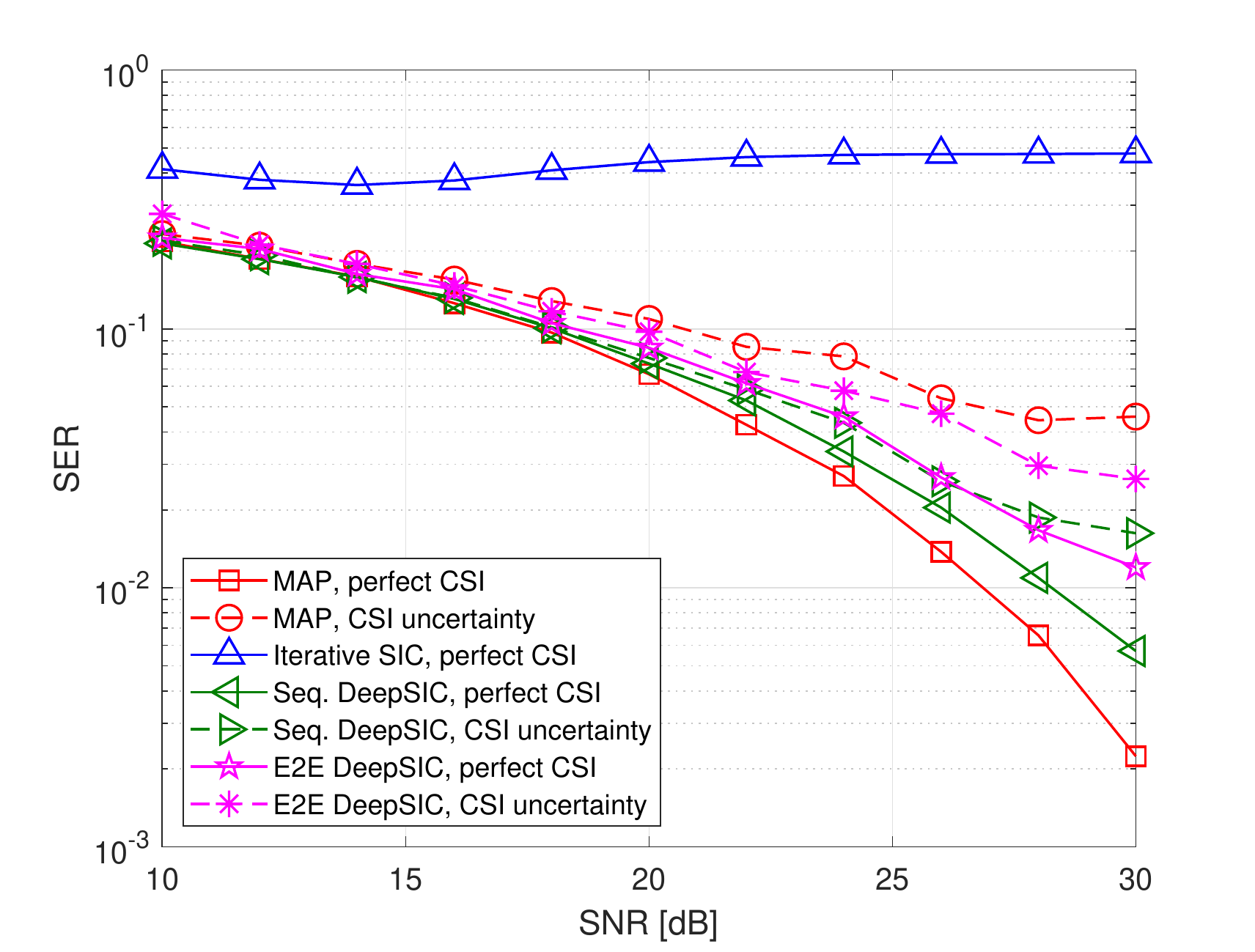}
	\figSpace
	\caption{\textcolor{NewColor}{\ac{ser} versus \ac{snr} of DeepSIC compared to the model-based \ac{map} rule and iterative \ac{sic} method, 
			$4\times4$ Poisson channel. For DeepSIC, {\em Perfect CSI} implies that the receiver is trained and tested using samples from the same channel, while under {\em CSI uncertainty} it is trained using samples from a set of different channels.}
	}
	\label{fig:Poisson4} 
\end{figure}

	The achievable \ac{ser} of Deep\ac{sic} versus \ac{snr} under both perfect \ac{csi} as well as \ac{csi} uncertainty with error variance $\SigE = 0.1$ is compared to the \ac{map} and iterative \ac{sic} detectors in Fig. \ref{fig:Poisson4}. Observing Fig. \ref{fig:Poisson4}, we again note that the performance of Deep\ac{sic} is only within a small gap of the \ac{map} performance with perfect \ac{csi}, and that the data-driven receiver is more robust to \ac{csi} uncertainty compared to the model-based \ac{map}. 
	In particular, Deep\ac{sic} with sequential training, which utilizes a deeper network architecture for each building block, outperforms here end-to-end training with basic two-layer structures for the conditional distribution estimation components. We conclude that under such non-Gaussian channels, more complex \ac{dnn} models are required to learn to cancel interference and carry out soft detection accurately. This further emphasizes the gain of our proposed sequential approach for training each building block separately, thus allowing to train an overall deep architecture using a limited training set based on the understanding of the role of each of its components. 
	 Furthermore, it is noted that iterative \ac{sic}, which is designed for linear Gaussian channels \eqref{eqn:Linear channel} in which the interference is additive, achieves very poor performance when the channel input-output relationship is substantially different from \eqref{eqn:Linear channel}. Since iterative \ac{sic} is shown to be unreliable in this setup with accurate knowledge of $\myMat{H}$, we do not include in Fig. \ref{fig:Poisson4} its \ac{ser} performance with \ac{csi} uncertainty.

	The results presented in this section demonstrate the ability of Deep\ac{sic} to achieve excellent performance and learn to implement interference cancellation from training, under statistical models for which conventional model-based interference cancellation is effectively inapplicable. 
	
%

	\vspace{-0.2cm}
	\subsection{Training Methods Comparison}
	\label{subsec:TrainSim}
	\vspace{-0.1cm}	
	In the numerical studies reported in the previous subsections, we used different network architectures for each training method of Deep\ac{sic}: For E2E Deep\ac{sic} we used two fully-connected layers in implementing the \acl{ml} based building blocks of Fig. \ref{fig:BasicModel2}, while Seq. Deep\ac{sic} used three layers. The rationale behind this setting was that for a given training set, Seq. Deep\ac{sic} updates distinct subsets of its parameters gradually and can thus adapt larger networks using the same training set compared to end-to-end training. However, it is also demonstrated in the previous subsections that E2E Deep\ac{sic} often achieves improved performance compared to Seq. Deep\ac{sic}, despite the fact that it uses smaller networks, as the joint training process results in a more accurate detector. 
	
\begin{figure}
	\centering
	\includegraphics[ width=\figWidth]{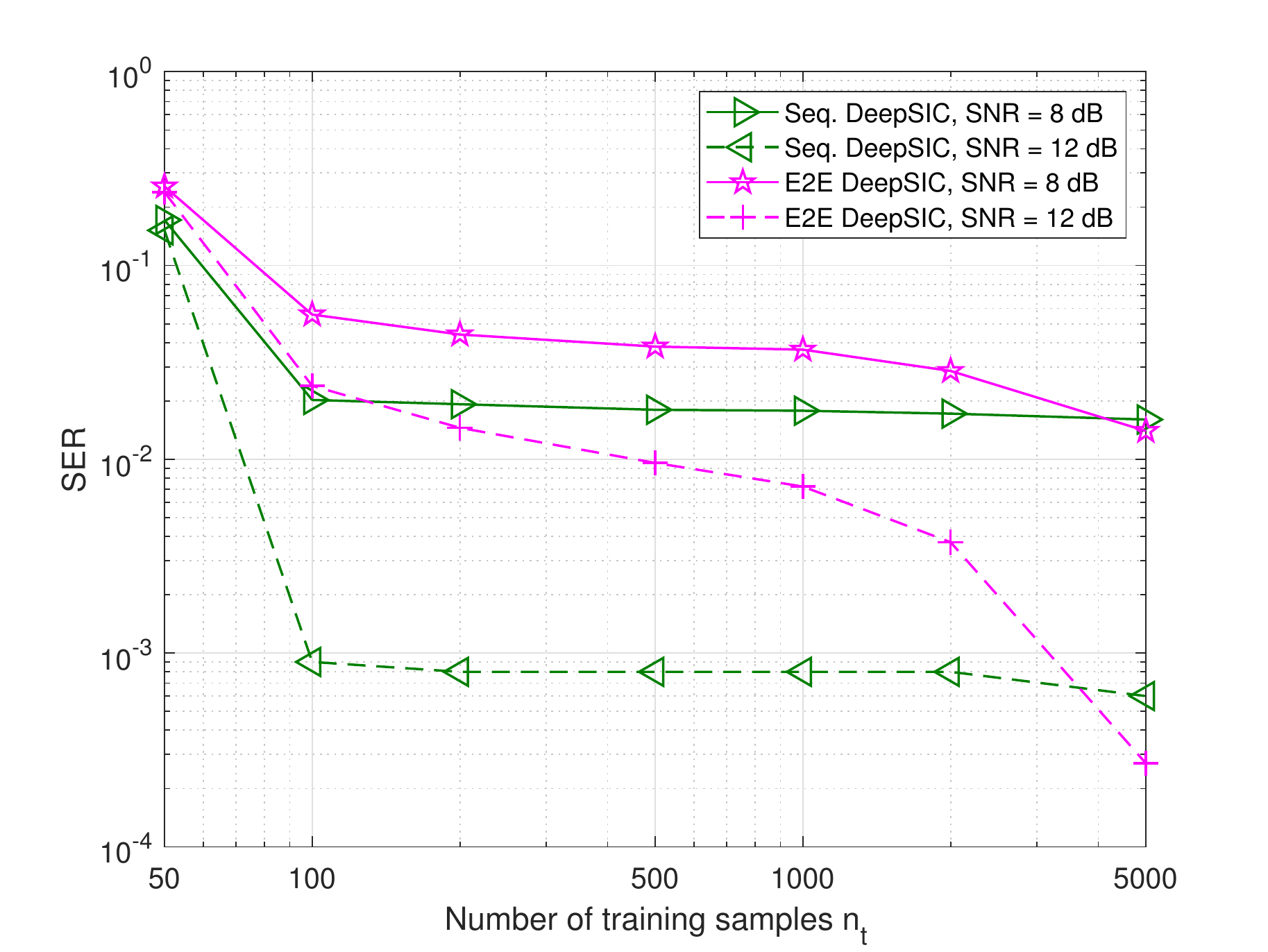}
	\figSpace
	\caption{\ac{ser} versus training set size of DeepSIC with sequential training and end-to-end training, $6\times 6$ linear channel with \ac{awgn}.
	}
	\label{fig:TrainignComp1} 
\end{figure}

	To understand which training method is preferable for a given network architecture and training set, in the following we numerically evaluate the performance of Deep\ac{sic} versus the training set size $\Ntraining$ when using the same number of parameters under both training methods. In particular, we fix the building blocks to consist of the architecture used under end-to-end training in the previous subsections, namely, a $(\Nantennas + (\Nusers - 1)(\CnstSize-1)) \times 60 \times \CnstSize$ fully-connected two layers. The achievable \ac{ser} of the training methods versus the training set size $\Ntraining$ for the $6 \times 6$ linear Gaussian channel  detailed in Subsection \ref{subsec:Linear} with perfect \ac{csi} for \ac{snr} values of $8$ and $12$ dB is depicted in Fig. \ref{fig:TrainignComp1}. 
	\textcolor{NewColor}{
	Observing Fig. \ref{fig:TrainignComp1}, we note that the accuracy of end-to-end training is substantially degraded as the number of labeled sampled decreases, since the available training is used to jointly adapt the complete set of parameters of the network. For comparison, sequential training, which exploits the understanding of the role of each building block in the model-based iterative \ac{sic} method to train each subsystem individually, is capable of operating reliably with as few as $\Ntraining =100$ training samples. In particular, it is observed in Fig. \ref{fig:TrainignComp1} that the fact that sequential training uses the same training set to adapt sub-sets of the overall architecture sequentially, facilitates convergence with much smaller training sets compared to end-to-end training. However, the fact that end-to-end training jointly adapts the complete DeepSIC architecture allows it to converge to improved configurations when a sufficient number of training samples is provided.	
	}
  These results demonstrate how converting a model-based algorithm into a data-driven system not only yields a suitable \ac{dnn} architecture, but can also facilitate its training. 

	\color{NewColor}
	\vspace{-0.2cm}
	\subsection{Online Tracking of Channel Variations}
	\label{subsec:Online}
	\vspace{-0.1cm}
	The ability of DeepSIC with sequential training to tune its parameters using small labeled data sets facilitates its application in dynamic time-varying channels. One of the major challenges of \acl{ml} aided receivers stems from the fact that \acp{dnn} require a large volume of training data in order to learn their mapping, and once trained they are applicable to inputs obeying the same (or a similar) distribution as the one used during training. Since communication channels are typically dynamic and conditions may change significantly over time, \ac{dnn}-based receivers should re-train periodically in order to track channel variations, without degrading the spectral efficiency. In such cases, the fact that DeepSIC can be efficiently trained with small data sets allows it to track time-varying channels without inducing additional communication overhead by exploiting the inherent structure of digital communication protocols. 
	
	We next demonstrate this advantage by applying the self-supervised online training method proposed in \cite[Sec. IV]{Shlezinger:19a} for tracking channel variations by utilizing the presence of coded communications. Here, each user transmits a set of codewords protected by a \ac{fec} code, and each codeword is transmitted as a block of symbols. The receiver applies DeepSIC to detect the transmitted symbols from the channel output and decodes the message using its \ac{fec} decoder. If the messages are successfully decoded, they are re-encoded and re-modulated into their corresponding channel input, which is used along with the observed channel output to re-train DeepSIC. Since the \ac{fec} decoder  can successfully recover the message even when some symbol errors are present, this approach allows an initially trained deep receiver which is capable of re-training using small data sets of the order of a single codeword, to track block-fading time-varying channel conditions.
	
\begin{figure}
		\centering
		\begin{subfigure}[b]{\linewidth}	
			\centering		
			\includegraphics[ width=\figWidth]{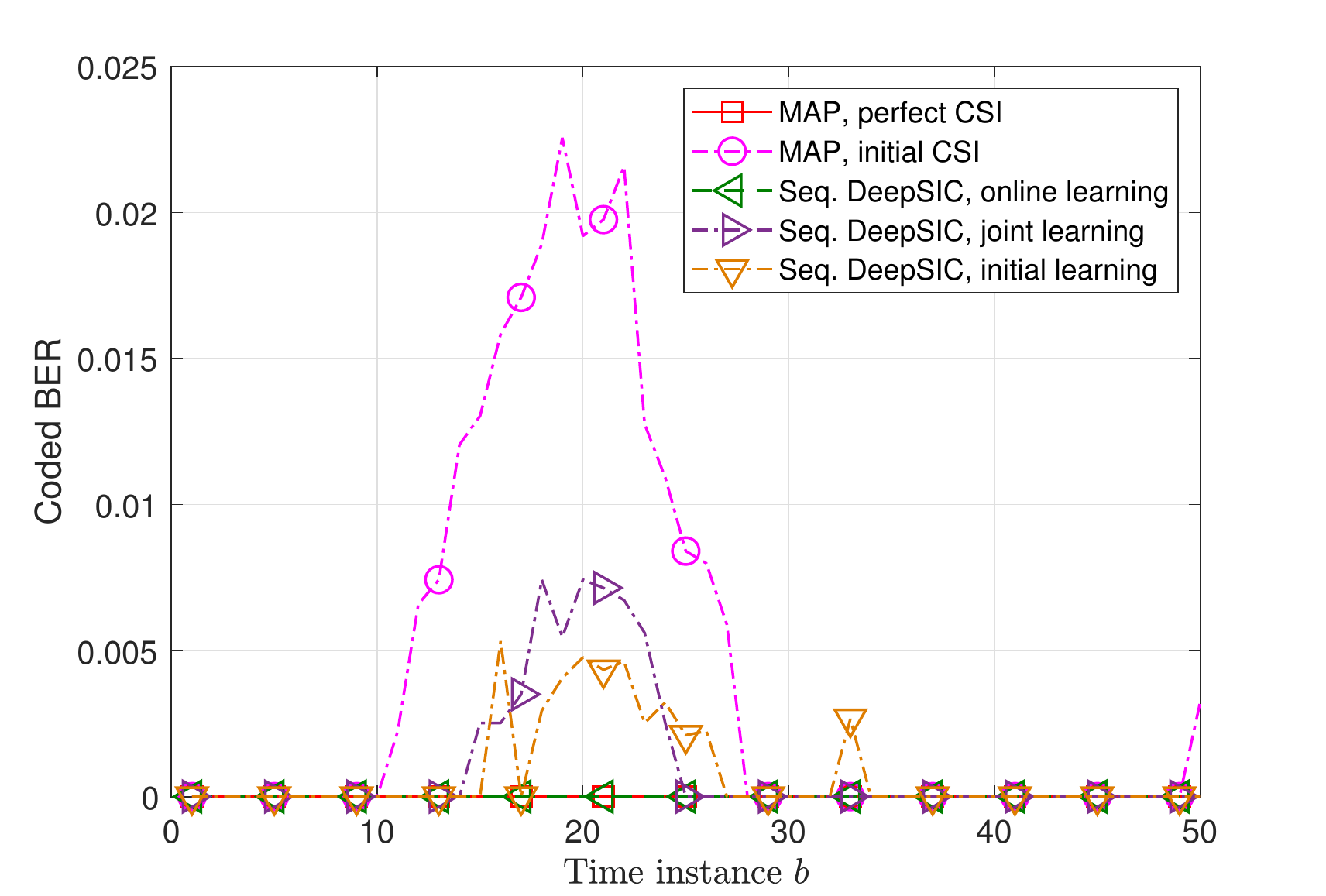}
			\figSpace
			\caption{Coded \ac{ber} for each codeword versus block index, linear channel with \ac{awgn}, ${\rm SNR} = 14$ dB.
			}	
			\label{fig:GaussianTV_SNR14} 
		\end{subfigure}
	
		\begin{subfigure}[b]{\linewidth}
			\centering			
			\includegraphics[ width=\figWidth]{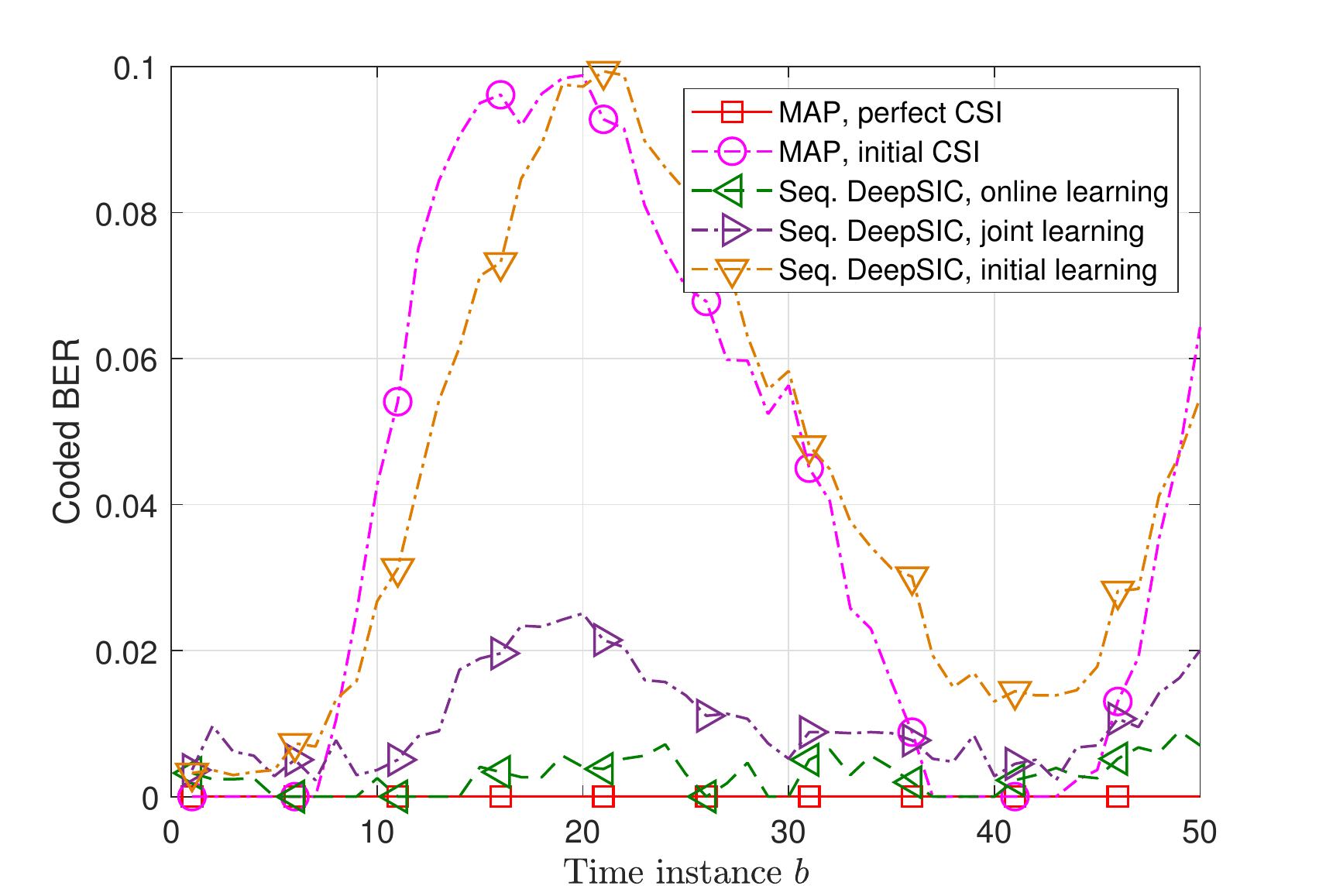}
			\figSpace
			\caption{Coded \ac{ber} for each codeword versus block index, Poisson channel, ${\rm SNR} = 32$ dB.
			}	
			\label{fig:PoissonTV_SNR32} 
		\end{subfigure}
	
		\begin{subfigure}[b]{\linewidth}	
			\centering		
			\includegraphics[ width=\figWidth]{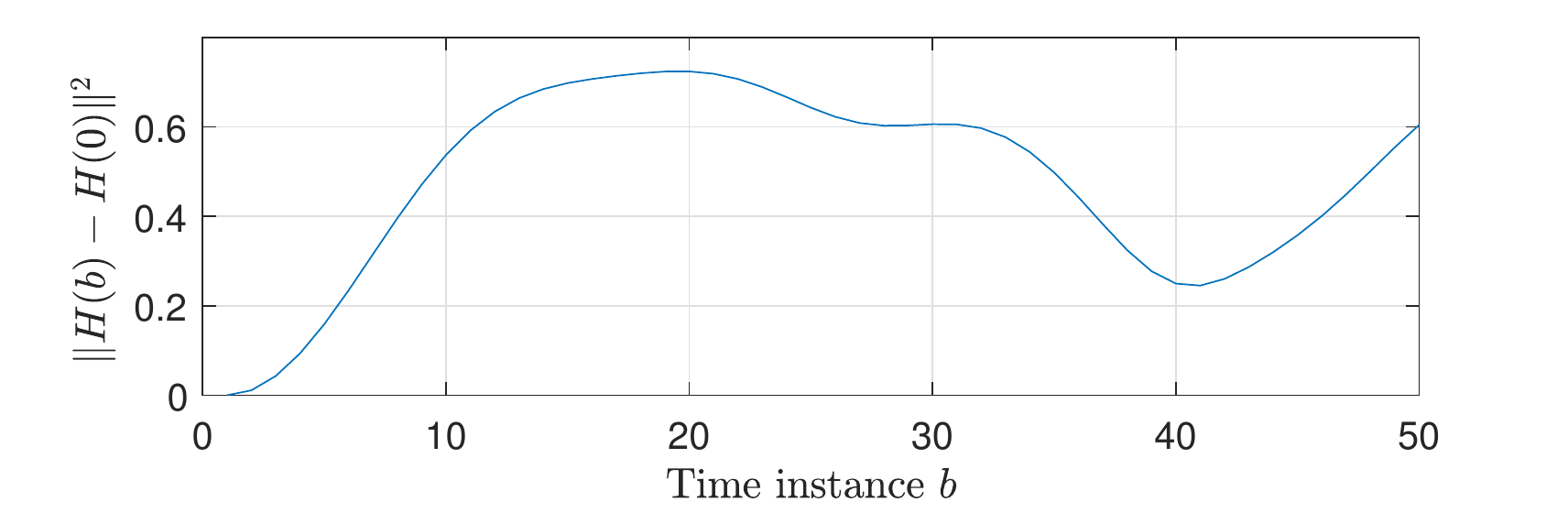}
			\figSpace
			\caption{Channel variations versus block index.
			}	
			\label{fig:ChannelTV_2} 
	\end{subfigure}
			\caption{Channel tracking via online self-supervised training.}	
			\label{fig:TVSim}
	\end{figure}

	In particular, we simulate two $4 \times 4$ channels: a linear \ac{awgn} channel \eqref{eqn:Linear channel} and a Poisson channel \eqref{eqn:PoissonChannel}. Here, each user transmits $50$ codewords encoded using a $[255,239]$ Reed-Solomon \ac{fec} code, i.e., each codeword consists of $2040$ symbols embedding a message of $1784$ bits. The channel observed by the $b$th codeword is generated with the channel matrix $\myMat{H}(b)$ whose entries are given by	
	\begin{equation}
	\left( \myMat{H}(b)\right)_{i,j} = e^{-|i-j|}\cos((\myVec{\phi})_i\cdot b). \qquad i \in \{1,\ldots, \Nantennas\}, j \in \{1,\ldots, \Nusers\},
	\label{eqn:ChannelMatTV}
	\end{equation}
	where we used $\myVec{\phi} = [51,39, 33, 21]$. The  difference between the initial channel matrix and that used during the $b$th block, given by the squared Frobenius distance $ \|\myMat{H}(b) - \myMat{H}(0)\|^2$, is depicted in Fig. \ref{fig:ChannelTV_2}. For each of the considered channel models, i.e., the linear \ac{awgn} channel and the Poisson channel, we evaluate the instantaneous \ac{ber} in decoding each codeword of DeepSIC using the method of \cite[Sec. IV]{Shlezinger:19a} for self-supervised channel tracking, where before the first block is transmitted, DeepSIC is trained using $5000$ samples corresponding to the channel with index $b=0$. These results are compared to the model-based \ac{map} detector which knows the channel at each time instance, as well as when it has only knowledge of the initial channel $\myMat{H}(0)$. Furthermore, we also evaluate the instantaneous \ac{ber} of DeepSIC without online training when trained only once using $5000$ samples corresponding to $\myMat{H}(0)$, as well as when the set of $5000$ training samples constitutes of $10$ subsets corresponding to $\myMat{H}(b)$ with $b \in \{1,\ldots,10\}$, representing joint  learning for multiple channels \cite{Park:20}. The results are depicted in Figs. \ref{fig:GaussianTV_SNR14}-\ref{fig:PoissonTV_SNR32} for the linear \ac{awgn} channel and the Poisson channel, respectively. We specifically consider relatively high \ac{snr} values, setting it to  $14$ db and $32$ dB for the Gaussian and Poisson channels, respectively, to focus on scenarios in which detection errors are expected to occur mostly due to channel variations, rather than due to the presence of a dominant noise, which is the case in lower \acp{snr}.
	
	Observing Figs.  \ref{fig:GaussianTV_SNR14}-\ref{fig:PoissonTV_SNR32}, we note that as the channel parameters begin to deviate considerably compared to their initial value, around the tenth codeword. Receivers which do not track channel variations, i.e., DeepSIC without retraining and the \ac{map} detector operating with the initial \ac{csi}, start exhibiting errors. However, DeepSIC which trains online in a self-supervised manner based on its \ac{fec} decoder outputs, successfully tracks the variations of the channel, demonstrating low error rates for each codeword, and in fact achieving zero errors over all considered blocks for the linear \ac{awgn} channel. This demonstrates how the ability of DeepSIC to adapt with a small number of samples using the sequential training method allows it to track time-varying channels without requiring additional pilots, by merely exploiting the inherent structure of digital communication protocols. Furthermore, we also observe that DeepSIC trained once over various channel conditions via joint learning exhibits reduced error rates compared to the \ac{map} detector with initial \ac{csi} as well as DeepSIC which trains only for the initial channel, particularly for the Poisson channel in Fig. \ref{fig:PoissonTV_SNR32}. This illustrates the potential of such joint training methods to improve the robustness of DeepSIC, as also observed for the \ac{csi} uncertainty case in the previous subsections.
	
	\color{black}

	\vspace{-0.25cm}
	\section{Conclusion}
	\label{sec:Conclusions}
	\vspace{-0.15cm}
 	In this work we proposed Deep\ac{sic}, a data-driven multiuser \ac{mimo} receiver architecture. To derive Deep\ac{sic}, we relied on the iterative \ac{sic} algorithm, which is an accurate and computationally feasible model-based \ac{mimo} detection scheme that can be naturally extended to incorporate \acl{ml} methods. We obtained the data-driven Deep\ac{sic} receiver by replacing the model-based building blocks of iterative \ac{sic} with dedicated compact \acp{dnn}. Unlike its model-based counterpart, Deep\ac{sic} is channel-model-independent and  can learn to implement interference cancellation in non-linear setups with non-additive interference. We proposed two methods for training Deep\ac{sic}: an end-to-end approach and a sequential scheme, where the latter is more suitable for small training sets and can thus be used to quickly adapt in dynamic environments. Our numerical results demonstrate that for conventional linear channels, Deep\ac{sic} approaches the \ac{map} performance, outperforming previously proposed \ac{dnn}-based receivers while demonstrating improved robustness to \ac{csi} uncertainty. Finally, we showed that the \ac{map}-comparable performance of deep\ac{sic} and its resiliency to uncertainty hold in the presence of non-linear channels, where the applicability of iterative \ac{sic} as well as previously proposed \acl{ml} based receivers is limited.
	

\end{document}

%% file: Preamble.tex
\newcommand{\myVec}[1]{{\boldsymbol{#1}}}
\newcommand{\myMat}[1]{{\mathsf{#1}}}
\newcommand{\mySet}[1]{\mathcal{#1}}
\newcommand{\myI}{{\myMat{I}}}			 		
\newcommand{\myY}{{\myVec{Y}}}
\newcommand{\myS}{{\myVec{S}}}

\newcommand{\Nusers}{K}
\newcommand{\Nantennas}{n_r}
\newcommand{\NusersSet}{\mySet{K}}
\newcommand{\Ntraining}{n_t}
\newcommand{\Nparameters}{P}
\newcommand{\SigW}{\sigma_w^2}
\newcommand{\SigE}{\sigma_e^2}
\newcommand{\Niter}{Q}
\newcommand{\NiterSet}{\mySet{Q}}
\newcommand{\Pdf}[1]{p_{ { #1}} }

\newcommand{\CnstSize}{M}			 			

\newcommand{\CovMat}[1]{\myMat{\Sigma}_{#1}}			

\newcommand{\Vecdim}[1]{} 

\definecolor{NewColor}{rgb}{0,0,0}

\acrodef{adc}[ADC]{analog-to-digital convertor}
\acrodef{cs}[CS]{compressed sensing}
\acrodef{dtft}[DTFT]{discrete-time Fourier transform}
\acrodef{dnn}[DNN]{deep neural network}
\acrodef{ml}[ML]{machine learning}
\acrodef{rnn}[RNN]{recurrent neural network}
\acrodef{csi}[CSI]{channel state information}
\acrodef{map}[MAP]{maximum a-posteriori probability}
\acrodef{snr}[SNR]{signal-to-noise ratio}
\acrodef{bs}[BS]{base station} 
\acrodef{iot}[IOT]{Interent of Things}
\acrodef{mimo}[MIMO]{multiple-input multiple-output}
\acrodef{mse}[MSE]{mean-squared error}
\acrodef{pdf}[PDF]{probability density function}
\acrodef{rv}[RV]{random variable}
\acrodef{fec}[FEC]{forward error correction}
\acrodef{rs}[RS]{Reed-Solomon}
\acrodef{lti}[LTI]{linear time-invariant}
\acrodef{wss}[WSS]{wide-sense stationary}
\acrodef{psd}[PSD]{power spectral density}
\acrodef{ser}[SER]{symbol error rate} 
\acrodef{ber}[BER]{bit error rate} 
\acrodef{isi}[ISI]{intersymbol interference} 
\acrodef{lstm}[LSTM]{long short-term memory} 
\acrodef{sgd}[SGD]{stochastic gradient descent} 
\acrodef{awgn}[AWGN]{additive white Gaussian noise}
\acrodef{bpsk}[BPSK]{binary phase shift keying}
\acrodef{sic}[SIC]{soft interference cancellation}
\acrodef{ofdm}[OFDM]{orthogonal frequency division multiplexing}

%% file: DeepSICArxiv_v06.bbl
\vspace{-0.25cm}
	\begin{thebibliography}{10}
		\vspace{-0.1cm}
		
		\bibitem{Shlezinger:20a}
		N. Shlezinger, R. Fu, and Y. C. Eldar.
		\newblock{``Deep soft interference cancellation for MIMO detection"}.
		\newblock{\em Proc. ICASSP}, Barcelona, Spain, May 2020.
		
		\bibitem{Foschini:98}
		G. J. Foschini and M. J. Gans.
		\newblock{``On limits of wireless communications in a fading environment when using multiple antennas"}.
		\newblock{\em Wireless Personal Communications}, vol. 6,  pp. 311--335, Mar. 1998.

		\bibitem{Marzetta:10}
		T. L. Marzetta.
		\newblock{``Noncooperative cellular wireless with unlimited numbers of base station antenna"}.
		\newblock{\em IEEE Trans. Wireless Commun.}, vol. 9, no. 11, Nov. 2010, pp. 3950--3600.		
		
		\bibitem{ElGamal:11}
		A. El Gamal and Y. H. Kim.
		\newblock{\em Network Information Theory}.
		\newblock Cambridge, 2011.			
	
		\bibitem{Shlezinger:17}
		N. Shlezinger and Y. C. Eldar.
		\newblock{``On the spectral efficiency of noncooperative uplink massive MIMO systems"}.
		\newblock {\em IEEE Trans. Commun.}, vol. 67, no. 3, Mar. 2019, pp. 1956--1971.	
			
		\bibitem{Andrews:05}
		J. G. Andrews.
		\newblock{``Interference cancellation for cellular systems: a contemporary overview"}.
		\newblock{\em IEEE Wireless Commun.}, vol. 12, no. 2, Apr. 2005, pp. 19--29.
							
		\bibitem{Choi:00}
		W. J. Choi, K. W. Cheong, and J. M. Cioffi.
		\newblock{``Iterative soft interference cancellation for multiple antenna systems"}.
		\newblock{\em Proc. WCNC}, Chicago, IL, Sep. 2000.		

		\bibitem{Wang:99}
		X. Wang and H. V. Poor.
		\newblock{``Iterative (turbo) soft interference cancellation and decoding for coded CDMA"}.
		\newblock{\em IEEE Trans. Commun.}, vol. 47, no. 7, Jul. 1999, pp. 1046--1061.

		\bibitem{Alexander:99}
		P. D. Alexander, M. C. Reed, J. A. Asenstorfer, and C. B. Schlegel.
		\newblock{``Iterative multiuser interference reduction: Turbo CDMA"}.
		\newblock{\em IEEE Trans. Commun.}, vol. 47, no. 7, Jul. 1999, pp. 1008--1014.
	
		\bibitem{Shlezinger:19b}
		N. Shlezinger, Y. C. Eldar and M. R. D. Rodrigues.
		\newblock{``Asymptotic task-based quantization with application to massive MIMO"}.
		\newblock{\em IEEE Trans. Signal Process.}  vol. 67, no. 15, Aug. 2019, pp. 3995-4012.
			
		\bibitem{Studer:16}
		C. Struder and G. Durisi.
		\newblock{``Quantized massive MU-MIMO-OFDM uplink"}.
		\newblock{\em IEEE Trans. Commun.}, vol. 64, no. 6, Jun. 2016, pp. 2387-2399.

		\bibitem{Iofedov:15}
		I. Iofedov and D. Wulich.
		\newblock{``MIMO-OFDM with nonlinear power amplifiers"}.
		\newblock{\em IEEE Trans. Commun.}, vol. 63, no. 12, Dec. 2015, pp. 4894-4904.


		\bibitem{Zheng:19}
		L. Zheng, M. Lops, Y. C. Eldar and X. Wang.
		\newblock{``Radar and communication coexistence: An overview: A review of recent methods"}.
		\newblock{\em IEEE Signal Process. Mag.}  vol. 36, no. 5, Sep. 2019, pp. 85-99.
 
		
		\bibitem{Khalighi:14}
		M. A. Khalighi and M. Uysal.
		\newblock{``Survey on free space optical Communication:
			A communication theory perspective"}.
		\newblock{\em IEEE Commun. Surv. Tut.}, vol. 16, no. 4,  2014, pp. 2231--2258.		

		\bibitem{Shlezinger:18b}
		N. Shlezinger, R. Shaked, and R. Dabora.
		\newblock{``On the capacity of MIMO broadband power line communications channels"}.
		\newblock{\em IEEE Trans. Commun.}, vol. 66, no. 10, Oct. 2018, pp. 4795-4810.	 
		
		
		\bibitem{Farsad:17}
		N. Farsad, Y. Murin, W. Guo, C-B. Chae, A. W. Eckford, and A. J. Goldsmith.
		\newblock{``Communication system design and analysis for asynchronous molecular timing channels"}.
		\newblock{\em IEEE Trans. Mol. Biol. Multi-Scale Commun.}, vol. 3, no. 4, Dec. 2017, pp. 239--253. 

		\bibitem{LeCun:15}
		Y. LeCun, Y. Bengio, and G. Hinton.
		\newblock{``Deep learning"}.
		\newblock{\em Nature}, vol. 521, no. 7553, May 2015, pp. 436--444.

		\bibitem{Oshea:17}
		T. O'Shea and J. Hoydis.
		\newblock{``An introduction to deep learning for the physical layer"}.
		\newblock{\em IEEE Trans. on Cogn. Commun. Netw.}, vol. 3, no. 4, Dec. 2017, pp. 563--575.
		
		\bibitem{Simeone:18}
		O. Simeone.
		\newblock{``A very brief introduction to machine learning with applications to communication systems"}.
		\newblock{\em IEEE Trans. on Cogn. Commun. Netw.}, vol. 4, no. 4, Dec. 2018, pp. 648--664. 	
		
		\bibitem{Mao:18}
		Q. Mao, F. Hu and Q. Hao.
		\newblock{``Deep learning for intelligent wireless networks: A comprehensive survey"}.
		\newblock{\em IEEE Commun. Surveys Tuts.}, vol. 20, no. 4,  2018, pp. 2595--2621.

		\bibitem{Gunduz:19}
		D. Gunduz, P. de Kerret, N. D. Sidiropoulos, D. Gesbert, C. Murthy, and M. van der Schaar.
		\newblock{``Machine learning in the air"}.
		\newblock{\em arXiv preprint}, arXiv:1904.12385, 2019.
		
		\bibitem{Balatsoukas:19}
		A. Balatsoukas-Stimming and C. Struder.
		\newblock{``Deep unfolding for communications systems: A survey and some new directions"}.
		\newblock{\em arXiv preprint}, arXiv:1906.05774, 2019.
		
		\bibitem{He:19c}
		H. He, S. Jin, C.-K. Wen, F. Gao, G. Y. Li, and Z. Xu. 
		\newblock{``Model-driven deep learning for physical layer communications"}.
		\newblock{\em arXiv preprint}, arXiv:1809.06059, 2018. 
	
		\bibitem{Bengio:09}
		Y. Bengio.
		\newblock{``Learning deep architectures for AI"}.
		\newblock{\em Foundations and Trends in Machine Learning}, vol. 2, no. 1, 2009, pp. 1--127.
	
		\bibitem{Kim:18}
		H. Kim, Y. Jiang, R. Rana, S. Kannan, S. Oh, and P. Viswanath.
		\newblock{``Communication Algorithms via deep learning"}.
		\newblock{\em arXiv preprint}, arXiv:1805.09317, 2018.
		
		\bibitem{Farsad:18}
		N. Farsad and A. Goldsmith.
		\newblock{``Neural network detection of data sequences in communication systems"}.
		\newblock{\em IEEE Trans. Signal Process.}, vol. 66, no. 21, Nov. 2018, pp. 5663--5678.
		
		\bibitem{Liao:19}
		Y. Liao, N. Farsad, N. Shlezinger, Y. C. Eldar, and A. J. Goldsmith.
		\newblock{``Deep symbol detection for millimeter wave communications"}.
		\newblock{\em Proc. GLOBECOM},  Waikola, HI, Dec. 2019.					 .
		
		\bibitem{Mosleh:18}
		S. Mosleh, L. Liu, C. Sahin, Y. R. Zheng, and Y. Yi.
		\newblock{``Brain-inspired wireless communications: Where reservoir computing
			meets MIMO-OFDM"}.
		\newblock{\em IEEE Trans. Neural Netw. Learn. Syst.},  vol. 29, no. 10, pp. 4694–4708, Oct 2018.
		
		\bibitem{Caciularu:18}
		A. Caciularu and D. Burshtein.
		\newblock{``Blind channel equalization using variational autoencoders"}.
		\newblock{\em arXiv preprint}, arXiv:1803.01526, 2018.
		
		\bibitem{Ye:18}
		H. Ye, G. Y. Li, and B. Juang.
		\newblock{``Power of deep learning for channel estimation and signal detection in OFDM systems"}.
		\newblock{\em IEEE Commun. Lett.}, vol. 7, no. 1, Feb. 2018, pp. 114-117.	
	
		\bibitem{Hershey:14}
		J. R. Hershey, J. Le Roux, and F. Weninger.
		\newblock{``Deep unfolding: Model-based inspiration of novel deep architectures"}.
		\newblock{\em arXiv preprint}, arXiv:1409.2574, 2014.	
		
		\bibitem{Gregor:10}
		K. Gregor and Y. LeCun.
		\newblock{``Learning fast approximations of sparse coding"}.
		\newblock{\em Proc. ICML}, Haifa, Israel, Jun. 2010.
		
		\bibitem{Solomon:18}
		O. Solomon, R. Cohen, Y. Zhang, Y, Yang, H. Qiong, J. Luo, R. J.G. van Sloun, and Y. C. Eldar.
		\newblock{``Deep unfolded robust PCA with application to clutter suppression in ultrasound"}.
		\newblock{\em arXiv preprint}, arXiv:1811.08252, 2018.		
		
		\bibitem{Li:19}
		Y. Li, M. Tofighi, J. Geng, V. Monga and Y. C. Eldar.
		\newblock{``An algorithm unrolling approach to deep blind image deblurring"}.
		\newblock{\em arXiv preprint}, arXiv:1902.03493, 2019.
		
		
		\bibitem{Wiesel:19}
		N. Samuel, T. Diskin and A. Wiesel.
		\newblock{``Learning to detect"}.
		\newblock{\em IEEE Trans. Signal Process.}, vol. 67, no. 10, May 2019, pp. 2554--2564.
		
		\bibitem{Corlay:18}
		V. Corlay, J. J. Boutros, P. Ciblat, and L. Brunel.
		\newblock{``Multilevel MIMO detection with deep learning"}.
		\newblock{\em arXiv preprint}, arXiv:1812.01571, 2018.		
		
		\bibitem{Takabe:19}
		S. Takabe, M. Imanishi, T. Wadayama, and K. Hayashi.
		\newblock{``Deep learning-aided projected gradient detector for massive
			overloaded MIMO channels"}.
		\newblock{\em arXiv preprint}, arXiv:1806.10827, 2018.	
		
		\bibitem{Khobahi:19}
		S. Khobahi, N. Naimipour, M. Soltanalian, and Y. C. Eldar.
		\newblock{``Deep signal recovery with one-bit quantization"}.
		\newblock{\em Proc. ICASSP}, Brighton, UK, May 2019.
	
			\bibitem{Liu:16}
		L. Liu, C. Yuen, Y. L. Guan, Y. Li, and Y. Su.
		\newblock{``Convergence analysis and assurance for Gaussian message passing iterative detector in massive MU-MIMO systems"}.
		\newblock{\em IEEE Trans. Wireless Commun.}, vol. 15, no. 9, Sep. 2016. pp. 6487--6501.		
		
		\bibitem{Liu:1b9}
		L. Liu, Y. Chi, C. Yuen, Y. L. Guan, and Y. Li.
		\newblock{``Capacity-achieving MIMO-NOMA: Iterative LMMSE detection"}.
		\newblock{\em IEEE Trans. Signal Process.}, vol. 67, no. 7, Apr. 2019, pp. 1758--1773.
		
		\bibitem{He:19a}
		H. He, C.-K. Wen, S. Jin, and G. Y. Li,
		\newblock{``A model-driven deep learning network for MIMO detection"}.
		\newblock{\em Proc. GlobalSIP}, Nov. 2018, pp. 584–588.
		
		\bibitem{He:19b}
		H. He, C.-K. Wen, S. Jin, and G. Y. Li,
		\newblock{`` Model-driven deep learning for	joint MIMO channel estimation and
			signal detection"}.
		\newblock{\em arXiv preprint}, arXiv:1907.09439, 2019.		

		\bibitem{Guo:19}
		J. Guo, B. Song, Y. Chi, L. Jayasinghe, C. Yuen, Y. L. Guan, X. Du, and M. Guizani.
		\newblock{``Deep neural network-aided Gaussian message passing detection for ultra-reliable low-latency communications"}.
		\newblock{\em Elsevier Future Generation Computer Systems}, vol. 95, Jun. 2019, pp. 629--638.	

		\bibitem{Zhang:19}
		Z. Zhang, Y. Li, C. Huang, Q. Guo, C. Yuen, and Y. L. Guan.
		\newblock{``DNN-aided block sparse Bayesian learning for user activity detection and channel estimation in grant-free non-orthogonal random access"}.
		\newblock{\em IEEE Trans. Veh. Technol.}, vol. 68, no. 12, Dec. 2019, pp. 12000--12011.		
				
		\bibitem{Shlezinger:19a}
		N. Shlezinger, N. Farsad, Y. C. Eldar, and A. J. Goldsmith.
		\newblock{``ViterbiNet:	A deep learning based Viterbi algorithm for symbol detection"}.
		\newblock{\em IEEE Trans. Wireless Commun.}, vol. 19, no. 5, May 2020. pp. 3319-3331.			
	
	
	\bibitem{Shlezinger:17b}
	N. Shlezinger, D. Zahavi, R. Murin, and R. Dabora.
	\newblock{``The secrecy capacity of Gaussian MIMO channels with finite memory"}.
	\newblock{\em IEEE Trans. Inform. Theory}, vol. 63, no. 3, Mar. 2017, pp. 1874-1897.
	
		\bibitem{Shlezinger:20b}
		N. Shlezinger, N. Farsad, Y. C. Eldar, and A. J. Goldsmith.
		\newblock{``Data-driven factor graphs for deep symbol detection"}.
		\newblock{\em Proc. ISIT}, Los Angeles, CA, Jun. 2020.	
	
	
		\bibitem{Shlezinger:20c}
		N. Shlezinger, N. Farsad, Y. C. Eldar, and A. J. Goldsmith.
		\newblock{``Inference from stationary time sequences via learned factor graphs"}.
		\newblock{\em arXiv preprint}, arXiv:2006.03258, 2020.	
							
		\bibitem{Varanasi:90}
		M. K. Varanasi and B. Aazhang.
		\newblock{``Multistage detection in asynchronous code-division multiple-access communications"}.
		\newblock{\em IEEE Trans. Commun.}, vol. 38, no. 4, Apr. 1990, pp. 505--519.		
				

		



		
		%
	

	
	

 



	


	

	

	
		\bibitem{Balevi:19}
		E. Balevi and J. G. Andrews.
		\newblock{``One-bit OFDM receivers via deep learning"}.
		\newblock{\em IEEE Trans. Commun.}, vol. 67, no. 6, Jun. 2019, pp. 4326-4336.	

		\bibitem{Shlezinger:19c}
		N. Shlezinger and Y. C. Eldar.
		\newblock{``Deep task-based quantization"}.
		\newblock{\em arXiv preprint}, arXiv:1908.06845, 2019.			
 

			




%
		

		

		






%




		
 
		
		
%
%
%
%
		

	
			

 
%
%
%
%
%
%
%


		

			

%
 
	
		 
%
	 


	 
%
%
%
%
%
 
 
 		\bibitem{Zhang:18}
 		Y. Zhang, T. Xiang, T. M. Hospedales, and H. Lu.
 		\newblock{``Deep mutual learning"}.
 		\newblock{\em Proc. CVPR}, Salt Lake City, UT, Jun. 2018.	 
 
  		\bibitem{Venkatakrishnan:13}
  		S. V. Venkatakrishnan, C. A. Bouman, and B. Wohlberg.
  		\newblock{``Plug-and-play priors for model based reconstruction"}.
  		\newblock{\em Proc. GlobalSIP}, Austin, Tx, Dec. 2013.
 
 
 
 		\bibitem{Dahlman:10}		
		 E. Dahlman, S. Parkvall, J. Skold, and P. Beming.
		 \newblock{\em 3G Evolution: HSPA and LTE for Mobile Broadband}.
		 \newblock Academic Press, 2010.
		 
		 
  
 		\bibitem{Wang:17}
		H. Wang, C. K. Wen, and S. Jin.
		\newblock{``Bayesian optimal data detector for mmwave OFDM system with low-resolution ADC"}.
		\newblock{\em IEEE J. Sel. Areas Commun.}, vol. 35, no. 9, Sep. 2017, pp. 1962--1979.
		
 		\bibitem{Park:20}
		 O. Simeone,  S. Park, and J. Kang.
		\newblock{``From learning to meta-learning: Reduced training overhead and complexity for communication systems"}.
		 \newblock{\em arXiv preprint}, arXiv:2001.01227, 2020.
		 


		%

		%

		%
		%
		
		
		
		
		%
		
		
	

		

		
		
	\end{thebibliography}
